\documentclass[twocolumn]{aastex6}

\usepackage{graphicx}
\usepackage{latexsym}
\usepackage{amsfonts,amsmath,amssymb}
\usepackage{epstopdf}
\usepackage{natbib}
\usepackage{mathrsfs}
\usepackage{algorithm}
\usepackage[utf8]{inputenc}
\usepackage{hyperref}
\usepackage[noend]{algpseudocode}

\def\apjref#1;#2;#3;#4 {\par\pni\ #1,  #2, {\bf #3}, #4. \par}

\newcommand{\beq}	{\begin{equation}}
\newcommand{\eeq}	{\end{equation}}
\newcommand{\beqa}{\begin{eqnarray}}
\newcommand{\eeqa}{\end{eqnarray}}
\newcommand{\avg}[1]  {{\langle #1 \rangle}} 

\newcommand{\e}	{$^{-1}$}

\def\simlt{\lower.5ex\hbox{$\; \buildrel < \over \sim \;$}}
\def\simgt{\lower.5ex\hbox{$\; \buildrel > \over \sim \;$}}

\def\vecnabla{
              \setbox1=\hbox{$\bigtriangledown$}
                           \raise.45ex\hbox{$\bigtriangledown$\hskip-.97\wd1
                           $\bigtriangledown$\hskip-.97\wd1
                           $\bigtriangledown$\hskip-.97\wd1}
                           \raise.47ex\hbox{$\bigtriangledown$}}

\def\rsun{\ifmmode {\rm R}_{\mathord\odot}\else $R_{\mathord\odot}$\fi}
\def\msun{\ifmmode {\rm M}_{\mathord\odot}\else $M_{\mathord\odot}$\fi}
\def\lsun{\ifmmode {\rm L}_{\mathord\odot}\else $L_{\mathord\odot}$\fi}

\newcommand{\mf}		{{m_f}}
\newcommand{\ml}		{{m_\ell}}

\newcommand{\scl}		{\Sigma_{\rm cl}}

\newcommand{\Ppt}		{\psi_{p2}}

\shortauthors{Gaches \& Offner}

\begin{document}
\title{Exploration of Cosmic Ray Acceleration in Protostellar Accretion Shocks and A Model for Ionization Rates in Embedded Protoclusters}

\author{Brandt A.L. Gaches}
\affil{Department of Astronomy, University of Massachusetts - Amherst}
\email{bgaches@astro.umass.edu}

\author{Stella S.R. Offner}
\affil{Department of Astronomy, The University of Texas at Austin}
\email{soffner@astro.as.texas.edu}

\begin{abstract}
We construct a model for cosmic ray acceleration from protostellar accretion shocks and calculate the resulting cosmic ray ionization rate within star-forming molecular clouds. We couple a protostar cluster model with an analytic accretion shock model to calculate the cosmic ray acceleration from protostellar surfaces. We present the cosmic ray flux spectrum from keV to GeV energies for a typical low-mass protostar. We find that at the shock surface the spectrum follows a power-law trend across 6 orders of magnitude in energy. After attenuation, the spectrum at high energies steepens, while at low energies it is relatively flat. We calculate the cosmic ray pressure and cosmic ray ionization rate from relativistic protons at the protostellar surface and at the edge of the core. We present the cosmic ray ionization rate for individual protostars as a function of their instantaneous mass and final mass. The protostellar cosmic ray ionization rate is $\zeta \approx 0.01 - 1$ s$^{-1}$ at the accretion shock surface. However, at the edge of the core, the cosmic ray ionization rate drops substantially to between $\zeta \approx 10^{-20}$ to $10^{-17}$ s$^{-1}$. There is a large spatial gradient in the cosmic ray ionization rate, such that inner regions may experience cosmic ray ionization rates larger than the often assumed fiducial rate, $\zeta = 3\times10^{-17}$ s$^{-1}$. Finally, we calculate the cosmic ray ionization rate for protostellar clusters over 5 orders of magnitude of cluster size. We find that clusters with more than approximately 200 protostars produce a higher cosmic ray ionization rate within their natal cloud than the fiducial galactic value.
\end{abstract}

\section{Introduction}
Cosmic rays (CRs) are one of the fundamental constituents of interstellar matter, along with ordinary matter, radiation and magnetic fields. Within molecular clouds, CRs are a primary driver of the complex chemistry in dense molecular gas \citep{grenier2015}. CRs, mostly relativistic protons, are the dominant source of ionization in molecular gas where ultraviolet radiation cannot penetrate.  At the temperatures and densities of molecular clouds, ion-neutral reactions make up the most efficient pathways \citep{watson1976, dalgarno2006}. Within molecular clouds, CR chemistry follows largely from the rapid formation of $H_3^+$:
\begin{equation*} {\rm
CR + H_2 \rightarrow H_2^+ + e^- + CR'
}
\end{equation*}
\begin{equation*}{\rm 
H_2^+ + H_2 \rightarrow H_3^+ + H,
}
\end{equation*}
where the CR' is the initial CR after the interaction. Following the formation of H$_3^+$, more complex molecules form via the generic reaction 
\begin{equation*}
{\rm X+H_3^+ \rightarrow HX^+ + H_2}
\end{equation*}
Observationally important molecules such as N$_2$H$^+$ and HCO$^+$ are created through this pathway. 

CRs are introduced into the chemistry through the CR ionization rate (CRIR), $\zeta$, which gives the rate of ionization per H ($\zeta$(H)) or per H$_2$ ($\zeta$(H$_2$)). In this work, we focus on $\zeta(H_2)$, which we hereafter refer to as $\zeta$. Observations of diffuse clouds find $\zeta \approx 10^{-16}$ s$^{-1}$ from measurements of both H$_3^+$ and H$_3$O$^+$ \citep{indriolo2007, indriolo2015}, and measurements near supernova remnants show even higher $\zeta$ \citep{indriolo2010}. Nearby supernova and winds from higher mass star are typically used to explain the CRIR in diffuse clouds \citep{amato2014}. Molecular clouds are expected to have a lower CRIR from energy losses due to gas interactions, and various screening mechanics are expected to reduce $\zeta$ with increasing gas column density \citep{padovani2009}.

Recent observational evidence has shown indirect evidence that the CRIR within protoplanetary disks (PPDs) and envelopes may be significantly greater than what would be expected with only Galactic CRs \citep{padovani2015, padovani2016}. It is not possible to detect CRs directly from embedded sources. Instead, the CRIR is inferred using various chemical signatures, often HCO$^+$ and N$_2$H$^+$. \cite{ceccarelli2014} used measurements of N$_2$H$^+$ and HCO$^+$ towards OMC-2 FIR 4, an intermediate mass protocluster and found $\zeta \approx 10^{-14}$ s$^{-1}$. \cite{podio2014} measured similar molecular ions towards the L1157-B1 shock, near the low-mass protostar L1157-mm, and found $\zeta = 3 \times 10^{-16}$ s$^{-1}$, which is inconsistent with the fiducial value from galactic sources if the CR flux is attenuated while penetrating into the cloud \citep{padovani2009}. The inferred spread and uncertainties in measured $\zeta$ are quite large \citep{favre2017}. 

In this work, we focus on the early stages of star formation when the protostar is still accreting much of its mass. \cite{padovani2013, padovani2014} show that the magnetic fields in dense cores can screen externally produced CRs. Furthermore, \cite{cleeves2013} studied 2D models of Class II PPDs and found that the T-Tauri wind was able to diminish the external CR flux by orders of magnitude. If CRs are screened in such a way, the higher values mentioned in the studies above indicate that locally accelerated CR may be important in star-forming regions. 

Within the solar system, there is also ample evidence that the young Sun produced high energy ($\ge$ 10 MeV) CRs. Measurements of short lived radio nuclei such as $^{10}$Be and $^{26}$Al indicate an over abundance in the early solar system \citep{gounelle2006a}. One possible explanation requires the interaction of dust particles with highly energetic CRs -- whether from galactic sources \citep{desch2004} or from the proto-Sun \citep{bricker2010, gounelle2013}. If we consider  the Sun a typical stellar object, it is likely many early stellar systems are bathed in highly energetic particles. In low-mass star-forming regions, like the Taurus Molecular Cloud, protostellar sources may be more important since there is a lack of supernova from the local star formation.

Theoretical studies of CR acceleration in protostars show that CR particles can be accelerated to MeV and GeV energies, both in their accretion shocks at the protostellar surface and within the jet shocks \citep{padovani2015, padovani2016}. For typical protostars, however, the unattenuated protostellar surface CR flux is a factor of 10$^4$ greater than the unattenuated flux produced by shocks associated with jets \citep{padovani2016}. Given that T Tauri stars exhibit enhanced stellar activity, \cite{rab2017} and \cite{rodgers-lee2017} adopted a scaled-up version of the solar spectrum and predicted a substantial increase in CR ionizations in protoplanetary disks.

A self-consistent treatment of the shock properties, which fully determine the CR spectrum and CRIR, and the CR physics is currently lacking. In the prior theory work, somewhat arbitrary assumptions are made about either the CR spectrum or the properties of the shock. In this work, we couple analytic models for protostar accretion histories and accretion shocks to produce self-consistent CR flux spectra for individual protostars and proto-clusters from high-energy protons accelerated at the protostellar surface.

We organize the paper as follows. In \S\ref{methods} we describe the analytic formalism of the protostars, the method for generating mock protostar clusters, and the CR physics. In \S\ref{results} we show the results of the model calculations and present CR spectra, pressures and ionization rates for individual protostars and the CRIR from protostellar clusters. In  \S\ref{disc} we discuss parameter variations and comparisons to observations. We summarize our results in \S\ref{summary}.

\section{Methods}\label{methods}

\subsection{Protostar Cluster Model}\label{cluster_model}

In this section we briefly summarize the Protostellar Mass Function (PMF) formalism of \cite{mckee2010} that we adopt. The PMF describes the underlying distribution of protostellar masses with the assumption of an accretion history, $\dot{m}$, and a final initial mass function (IMF), $\Psi$. We assume a truncated Chabrier IMF \citep{chabrier2005}, where we denote the upper truncation mass $m_u$. The bi-variate number density of protostars within a cluster is
\beq
d^2N_p = N_p \Ppt(m, \mf) d\ln m d \ln \mf,
\eeq
where $N_p$ is the number of protostars in the cluster, $\Ppt$ is the bi-variate PMF, $m$ is a protostar's current mass and $\mf$ is the expected final mass. \cite{mckee2010} showed that for a steady star formation rate
\beq
\Ppt(m, \mf) = \frac{m\Psi(\mf)}{\dot{m}\avg{t_f}},
\eeq
where $t_f$ is the time it takes to form a star with mass $\mf$ and 
\beq
\avg{t_f} = \int\limits_{\ml}^{m_u} d\ln \mf \Psi(\mf) t_f(\mf).
\eeq
Following \cite{gaches2018}, we adopt the Tapered Turbulent Core (TTC) accretion history \citep{mckee2003, offner2011}:
\beq\label{eq:mdot}
\dot m_{\rm TTC}=\dot{m}_{\rm TC} \left(\frac{m}{\mf}\right)^{1/2} \mf^{3/4}  \left [ 1 - \left (\frac{m}{m_f} \right )^{1/2} \right ]^{1/2}
~~~M_\odot~\mbox{yr\e},
\eeq
This model produces higher accretion rates for higher mass stars and smaller accretion rates as protostars approach their final mass. \cite{mckee2003}
adopt 
\beq\label{eq:mdotnorm}
\dot{m}_{\rm TC} = 3.6\times 10^{-5}\scl^{3/4} ~~~M_\odot~\mbox{yr\e}
\eeq
where $\scl$ is the surface density given in units of g cm$^{-2}$ for a star-forming clump. The formation time $t_f$ is 
\beq
t_f = \frac{4}{\dot{m}_{\rm TC}} \mf^{1/4}.
\eeq

\subsection{Cluster Generation and Statistical Sampling}\label{sampling}

We model clusters with different sizes and star formation efficiencies  following \cite{gaches2018}. Given a cluster with $N_*$ protostars, the total mass is well-approximated by $M_* \approx \avg{m} N_*$. We denote the efficiency, $\epsilon_g = \frac{M_*}{M_{\rm gas}}$. We approximate a cloud as a uniform density sphere with radius, $R = \left ( \frac{3 M_{\rm gas}}{4 \pi \rho} \right )^{1/3}$, where $\rho = \mu_{\rm M} m_{\rm H} n$, $n$ is the gas number density, and $\mu_{\rm M}$ is the mean molecular weight for cold molecular gas. The gas surface density is $\scl = \frac{M_g}{\pi R^2}$, which sets $\dot{m}_{\rm TTC}$ for the cluster.

We generate mock clusters following the method in \cite{gaches2018}. We directly draw $N_*$ ($m$, $\mf$) pairs from the bi-variate PMF using the conditional probability method. First, we marginalize $\Ppt$ over the final mass, $\mf$, yielding $\Psi(m)$. The one-dimensional distribution is sampled using the inversion method. We use the $m$ samples to calculate the one-dimensional conditional probability: $\Psi(\mf | m) = \frac{\Ppt(m = m, \mf)}{\Psi(m = m)}$. In this work, we generate $N_{\rm cl}$, mock clusters when calculating cluster-wide statistics (such as the total cluster CRIR) to reduce statistical noise.

\subsection{Accretion Shock Model}\label{shock_model}

The protostellar accretion shock occurs  at the protostellar surface, $r_*$. The shock front is assumed to be stationary, and the shock velocity is taken to be the Keplerian velocity
\beq
\label{eq:vs}
v_{\rm s} = \sqrt{\frac{2Gm}{r_*}} = 309 \, \left (\frac{m}{0.5 M_{\odot}} \right )^{0.5} \left (\frac{r_*}{2 R_{\odot}} \right )^{-0.5} {\rm \, \, km \, s^{-1}}
\eeq
where $r_*$ is the protostellar radius calculated using the model presented in \cite{offner2011}. In the strong shock regime, the shock temperature is
\beq
\label{eq:ts}
T_s = \frac{3}{16}\frac{\mu_{\rm I} m_{\rm H}}{k} v_s^2 = 1.302\times10^6 \, \left ( \frac{\mu_{\rm I}}{0.6} \right ) \left ( \frac{v_s}{309 {\rm \, km \, s^{-1}}} \right )^2 {\rm \, K}
\eeq
where $\mu_{\rm I}$ is the mean molecular weight for ionized gas. The accretion onto the protostar is thought to occur in columns following the magnetic field lines \citep{hartmann2016}. Within these flows, the shock can be treated as planar and vertical, such that the shock front normal is parallel to the field lines. The density of the accreted material is given by the accretion rate and the filling fraction of the accretion columns on the surface of the protostar. The shock density is then
\begin{eqnarray}\nonumber\label{eq:rhos}
\rho_s &=& \frac{\dot{m}}{A v_s} = 8.387\times10^{-10} \left ( \frac{\dot{m}}{10^{-5}  {\rm \, M_{\odot} \, yr^{-1}}} \right ) \left ( \frac{f}{0.1} \right )^{-1} \times \\
&& \left ( \frac{r_*}{2 {\rm \, R_{\odot}}} \right )^{-2} \left ( \frac{v_s}{309 {\rm \, km \, s^{-1}}} \right )^{-1} {\rm \, \, g \, cm^{-3}}
\end{eqnarray}
where $A$ is the area of the accretion columns, $A = 4 \pi f r_*^2$, and $f$ is the filling fraction. We adopt a constant value of $f = 0.1$, which reflects the high accretion rates typical of protostars. However, we note that the filling fraction likely depends on accretion rate and time, and thus, $f>0.1$ for very young protostars and declines during the protostellar phase \citep{hartmann2016}. The number density of the shock is $n_s = \frac{\rho_s}{\mu_{\rm I} m_{\rm H}}$, where we assume the gas is fully ionized fully \citep{hartmann2016},  and we use $\mu_{\rm I} = 0.6$ for a fully ionized gas.
\subsection{Cosmic Ray Model}\label{CR_model}

\subsubsection{Cosmic Ray Spectrum}
The physics of CR acceleration has been relatively well understood for decades \citep[i.e.,][]{umebayashi1981, drury1983}. First-order Fermi acceleration, also known as Diffusive Shock Acceleration (DSA), can work in jet shocks and protostellar accretion shocks to produce high-energy CRs \citep{padovani2016}. Under this mechanism, CR protons gain energy every time they pass across the shock. If the flow is turbulent, magnetic field fluctuations scatter these protons back and forth across the shock many times allowing them to continuously gain energy. However, several important conditions must be met for DSA to occur. The flow must be supersonic and super-Alfv\'{e}nic for there to be sufficient magnetic fluctuations. The acceleration rate must be greater than the collisional loss rate, the wave dampening rate, and the rate of diffusion in the transverse direction of the shock. Finally, the acceleration time must be shorter than the timescale of the shock. Each of these timescale conditions limits the energy at which the CRs can be accelerated (as discussed below and in Appendix \ref{CRphysics}). 
We verify that all of these conditions are met throughout our parameter space \citep[see also][]{padovani2016}. 

We describe in detail the relevant physics for accretion shocks in Appendix \ref{CRphysics} following \cite{padovani2016}. Here we give a brief summary of the model. First-order Fermi acceleration leads to a power-law momentum distribution, $f(p)$, where
\beq\label{eq:fp}
f(p) \propto p^{-q}.
\eeq 
The physical quantity of most interest in this work is the CR flux spectrum, $j(E)$. The flux spectrum is related to the accelerated number density spectrum, $\mathcal{N}(E)$ by:
\beq
j(E) = \frac{v(E) \mathcal{N}(E)}{4\pi} ~~ ({\rm particles ~GeV^{-1} ~cm^{-2} ~s^{-1} ~sr^{-1}})
\eeq
where $v(E)$ is the velocity as a function of energy in the the relativistic limit. The number density spectrum is related to the more fundamental momentum distribution
\beq
\mathcal{N}(E) = 4\pi p^2 f(p) \frac{dp}{dE} ~~ ({\rm particles ~GeV^{-1} ~cm^{-3}}),
\eeq
where the momentum distribution power-law index, $q$, depends on the underlying shock properties. The flux spectrum is defined between an energy range, $E_{\rm inj} < E < E_{\rm max}$, where E$_{\rm inj}$ is the injection energy scale of thermal CR particles and E$_{\rm max}$ is the maximum energy possible for acceleration to be efficient. E$_{\rm inj}$ depends on the strength of the shock and the hydrodynamic properties, such that stronger shocks and stiffer equations of state lead to an injection energy increase. E$_{\rm max}$ is determined by a combination of the magnetic field, the ionization fraction, and the shock acceleration efficiency. It is important to note that as long as E$_{\rm max} > 1$ GeV, any additional increases only weakly affects our results below. We are mainly interested in the effects of ionization produced by CRs, which is dominated by CRs with energies between 100 MeV and 1 GeV \citep{indriolo2010}. Higher energy CRs are important to understand and characterize gamma rays or similar high-energy phenomena. 

Neutral gas is not only ionized by the primary CRs. Electrons produced by CR ionization can have sufficient energy to cause additional ionizations.  We account for secondary electron ionizations following \cite{ivlev2015}, which we discuss in detail in Appendix \ref{ap:secelectrons}. We ignore the effects of primary electron acceleration. Electrons couple more strongly to the magnetic field and have a significantly lower mass than protons resulting in a maximum energy orders of magnitude below that of protons \citep{padovani2016}.

The most uncertain parameters are the magnetic field, $B$, and the shock efficiency parameter, $\eta$. The latter represents the fraction of particles accelerated from the thermal population. Before selecting the fiducial values, we explore the impact of each on the results. The magnetic field at the surface of a protostar has not been accurately measured. Theoretical studies of Class II/T-Tauri stars predict a surface magnetic field of a few Gauss to 1 kG \citep{johns-krull2007}.  We vary $\eta$ between 10$^{-5}$ and 10$^{-3}$ and find that the CRIR scales linearly with $\eta$ and changes $E_{\rm max}$ by factors of a few. However, a value of $\eta = 10^{-3}$ is an extreme case. We fix $B = 10$ G and $\eta = 10^{-5}$ following \cite{padovani2016}. Table \ref{tab:params} summarizes our fiducial physical parameters. We discuss the effects of varying these parameters in \S\ref{disc}.

\begin{deluxetable}{ccc}
\tablecolumns{3}
\tablecaption{Model Parameters \label{tab:params}}
\tablehead{\colhead{Parameter} & \colhead{Fiducial Value} & \colhead{Range}}
\startdata
$\mu_{\rm I}$ & 0.6 (ionized) & \\ 
$\mu_{\rm M}$ &  2.8 (neutral molecular) & \\
$\rho$ (\S\ref{sampling})& $10^3\mu$m$_H$  cm$^{-3}$ & \\ 
$\Sigma_{\rm cl}$ & 1.0 g cm$^{-2}$ & \\
f & 0.1 & 0.1 - 0.9 \\
B & 10 G & 10 G - 1 kG \\
$\eta$ & 10$^{-5}$ & 10$^{-5}$ - 10$^{-3}$ \\
\enddata
\tablecomments{Values for the parameters we assume in the model, and the ranges we discuss in \S\ref{disc}}
\end{deluxetable}

\subsubsection{Cosmic Ray Interactions and Ionization Rate}\label{CRIR_model}

The CRIR from protons and secondary electrons as a function of gas column density is 
\beq
\zeta(N) = 2\pi \int \left [ j(E,N) \sigma_p^{\rm ion}(E) + j_e^{\rm sec}(E, N) \sigma_e^{\rm ion}(E) \right ] dE, 
\eeq
where $j(E, N)$ is the CR flux at energy E after traveling through the column density, $N$, and $\sigma_k^{\rm ion}$ is the ionization cross section \citep{padovani2009}.  \cite{krause2015} proposed relativistic corrections to the $p-{\rm H_2}$ ionization cross sections, applicable when $E > 1$ GeV. We confirmed that the correction factor to the cross section has no impact on our results due to the small population of GeV CRs. Therefore, we use the non-relativistic cross section for simplicity. At the shock, the CRIR is expected to be considerable. However, as CRs propagate away from the protostars they undergo two different processes: energy losses due to collisions with matter and geometric dilution. The former directly modifies the spectrum's shape, since the energy loss is not a grey process (with respect to energy). The latter reduces the overall flux. We use the formalism of \cite{padovani2009} to account for the energy losses from interactions with matter. The loss function is defined by
\beq
L(E) = -\frac{dE}{dN({\rm H_2})}.
\eeq
We can calculate the new energy, E$_k$, after losses as a function of the initial energy, E$_0$ for a specific column density, $N({\rm H_2})$:
\beq
N({\rm H_2}) = n({\rm H_2}) \left [ R(E_0) - R(E_k) \right ] 
\eeq
with the range, $R(E)$, defined as
\beq
R(E) = \frac{1}{n({\rm H_2})} \int_0^{E} \frac{dE}{L(E)}.
\eeq
The attenuated spectrum is calculated assuming the number of particles is conserved:
\beq
j'(E_k, N({\rm H_2})) = j(E_k, N=0) \frac{L(E_0)}{L(E_k)} \label{jatten}
\eeq
Equation \ref{jatten} only takes into account interactions with matter. However, the CRs are generated by a point source, so we must also take into account the spatial dilution of the flux. This is in contrast to \cite{padovani2009} who consider a plane parallel slab geometry. We account for the spatial dilution by modifying the attenuated flux as:
\beq
j(E_k, N({\rm H_2})) = j'(E_k, N({\rm H_2})) \left ( \frac{R_*}{(R_* + r(N))}\right )^a,
\eeq
where $r(N({\rm H_2})$ is the radius at which the gas has column density $N({\rm H_2})$, and $a$ is the power-law index for how fast the flux is diluted. A full solution of the CR transport equation is needed to properly determined $a$. However, we take $a = 2$, corresponding to free streaming, as a lower limit for the CRIR, which is a common assumption \citep[e.g.][]{turner2009, rab2017}. Observations of ions in protostellar envelopes may be able to constrain the transport further, primarily whether they undergo free streaming ($a = 2$) or diffusive ($a = 1$) transport. We discuss the implications of different transport regimes in \S\ref{sec:transport}.

The H$_2$ column density, $N({\rm H_2})$, from the embedded protostar to the edge of the core is the final piece needed to relate the protostellar feedback to the natal cloud environment. We use the \cite{mckee2003} model describing protostellar accretion from a turbulent core to calculate appropriate column densities. For a dense core embedded in a turbulent star-forming clump, the envelope column density and core radius are given by:
\begin{align}\label{eq:sigcore}
\Sigma_{\rm core} &= 1.22 \Sigma_{\rm cl} \\
N(H_2)_{\rm core} &= \frac{\Sigma_{\rm core}}{\mu_{\rm M} m_{\rm H}}
\end{align}
\beq\label{eq:rcore}
R_{\rm core} = 0.057 \Sigma_{cl}^{-\frac{1}{2}} \left ( \frac{m_f}{30 \, {\rm M_{\odot}}}\right )^{\frac{1}{2}} \, \, {\rm pc},
\eeq
where $\Sigma_{\rm cl}$ is the surface density of the embedding clump, which is the normalization factor in $\dot{m}_{\rm TTC}$, and $N({\rm H_2})$ is the H$_2$ column density. 

Our cluster results do not depend on an assumed density profile. However, we adopt a density profile to calculate how the CRIR changes within a protostellar enveloped (See \S\ref{CRIRsingle}). We calculate the radius for a given column density by assuming a polynomial density distribution, $n(r) = n_s \left ( \frac{R_{\rm core}}{r} \right )^{-\kappa_\rho}$, where $n_s$ is the number density at the surface of the core and $\kappa_\rho = \frac{3}{2}$ is motivated by \cite{mckee2003}. The column density as measured from the protostar follows from $N_{\rm H_2}(r) = N({\rm H_2})_{\rm core} - \int_r^{R_{\rm core}} n(r) dr$.  Inversion results in $r(N({\rm H_2})) = \left ( \frac{2 n_s R_{\rm core}^{\frac{3}{2}}}{N({\rm H_2)}_{\rm core} - 2n_s R_{\rm core} - N({\rm H_2}) }\right )^2$.

The total CRIR produced by a forming star cluster is calculated by:
\beq
\zeta(N_*) = \sum_i^{N_*} \int  \left [ j_i(E,N_i) \sigma_p^{\rm ion}(E) + j_{i,e}^{\rm sec}(E, N_i) \sigma_e^{\rm ion}(E) \right ]  \, dE, 
\eeq
where $N_i$ is the H$_2$ column density from the protostar to the surface of the core. 

\section{Results}\label{results}

\subsection{Dependence on Protostellar Mass}
\subsubsection{Flux Spectrum}\label{specsec}

The CR spectrum of protostars is an observational unknown. The unattenuated spectrum is impossible to constrain because CRs quickly interact with matter, both neutral, in the form of excitations and ionizations, and ionized, through Coulomb interactions. Since protostars are embedded within their natal envelope, their radiation is heavily re-processed by the surrounding dust. Current observations cannot differentiate between the CRs accelerated by Galactic sources and the Sun versus protostellar sources. Previous studies of young stellar objects have therefore used scaled versions of the local solar spectrum \citep{rab2017, rodgers-lee2017}. In this section, we present predictions for the CR flux spectrum both at the protostellar surface and at the edge of its core.

\begin{figure}
\plotone{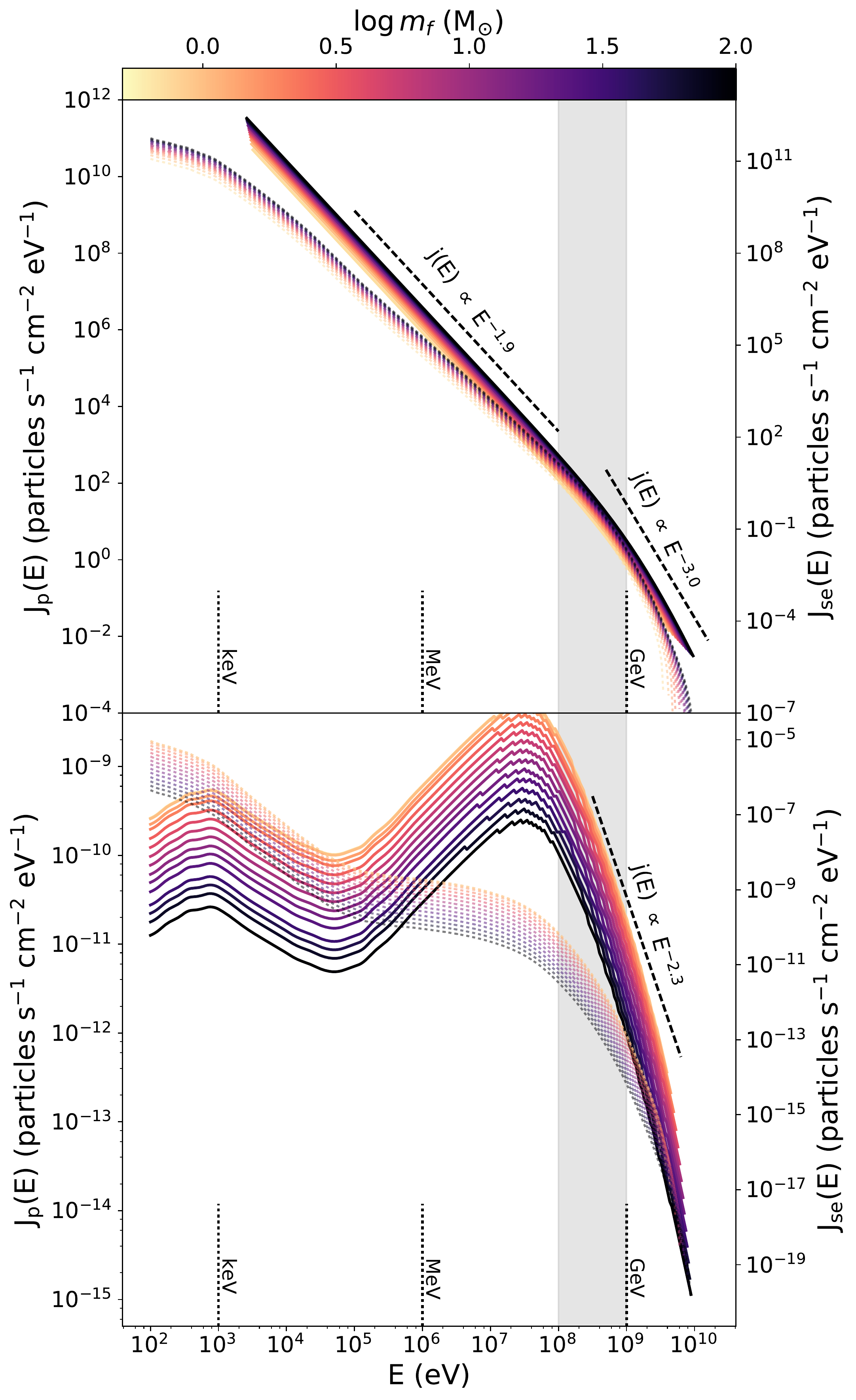}
\caption{\label{fig:spec} Proton (solid) and secondary electron (dotted) CR flux spectrum as a function of energy for a $m = 0.5$ M$_{\odot}$ protostar.  The color indicates the final mass, $\mf$, of the protostar. The vertical grey band shows the dominant energy range for ionization. Power-law fits to various parts of the spectra are presented as the dashed lines and annotations. Top: Unattenuated flux at the protostellar accretion shock surface. Bottom: Attenuated flux at the edge of the core. We set $\Sigma_{\rm cl} = 1.0$ g cm$^{-2}$.} 
\end{figure}

Figure \ref{fig:spec} shows the CR spectrum generated by the accretion shock for a protostar with an instantaneous mass $m = 0.5$ M$_{\odot}$ as a function of its final mass taking into account both protons and secondary electrons assuming $\Sigma_{\rm cl} = 1.0$ g cm$^{-2}$. The unattenuated spectrum shows a clear power-law behavior with an index of -1.9. In the strong shock regime, the index in Equation \ref{eq:fp} asymptotically approaches $q = 4$. The energy spectrum scales as $p^2 f(p)$, such that $j(E) \propto p^{-2}$ \citep{amato2014},  which is consistent with our result. The unattenuated spectrum is well described as a product of an efficient strong shock at the protostellar surface. We note that the final mass dependence acts largely to scale the spectrum through the accretion rate. 

The injection energy of 1 keV corresponds to an ionized plasma with a temperature of roughly $1.5 \times 10^6$ K. The proton spectrum shows that the maximum energy weakly scales with the final mass of the protostar (or the accretion rate). The spectrum tapers to $q = 3.0$ at high energies due to the acceleration inefficiency at such relativistic speeds. The energy corresponding to the turnover for all spectra, $E \approx 1$ GeV, is the transition where $E = m_p c^2$. The secondary electron spectrum likewise shows qualitatively similar behavior.

The attenuated spectrum in Figure \ref{fig:spec} shows very different behavior. Interactions with the dense core greatly alter the shape of the spectrum, and the radial distance traveled significantly reduces the flux. Short-ward of 1 GeV, ionizations and excitations effectively flatten the spectrum and shift higher energy CRs to lower energies. Losses due to pions are minimal due to the lack of CRs above 10 GeV. From 100 MeV to 1 GeV the proton flux spectrum exhibits a power-law index of $q = 2.5$. 

The secondary electron spectrum shows similar features from collisional losses. However, the interactions of higher energy CRs enhances the secondary electron spectrum such that there are significantly more lower energy electrons. At $E = 1$ keV for every CR proton there are 10$^4$ secondary electrons due to the interactions of higher energy CRs, which are less affected by collisional losses. 

\begin{figure*}
\plottwo{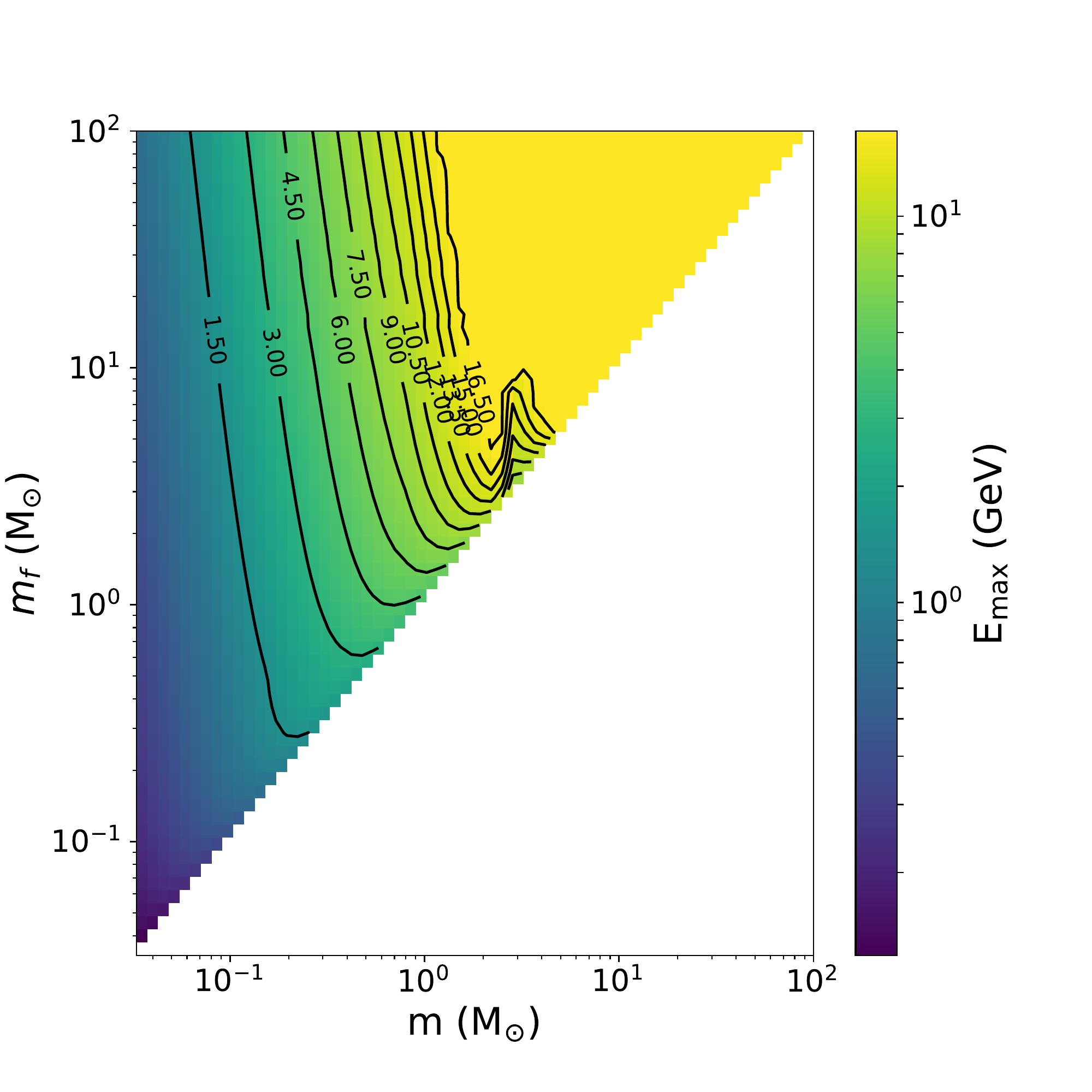}{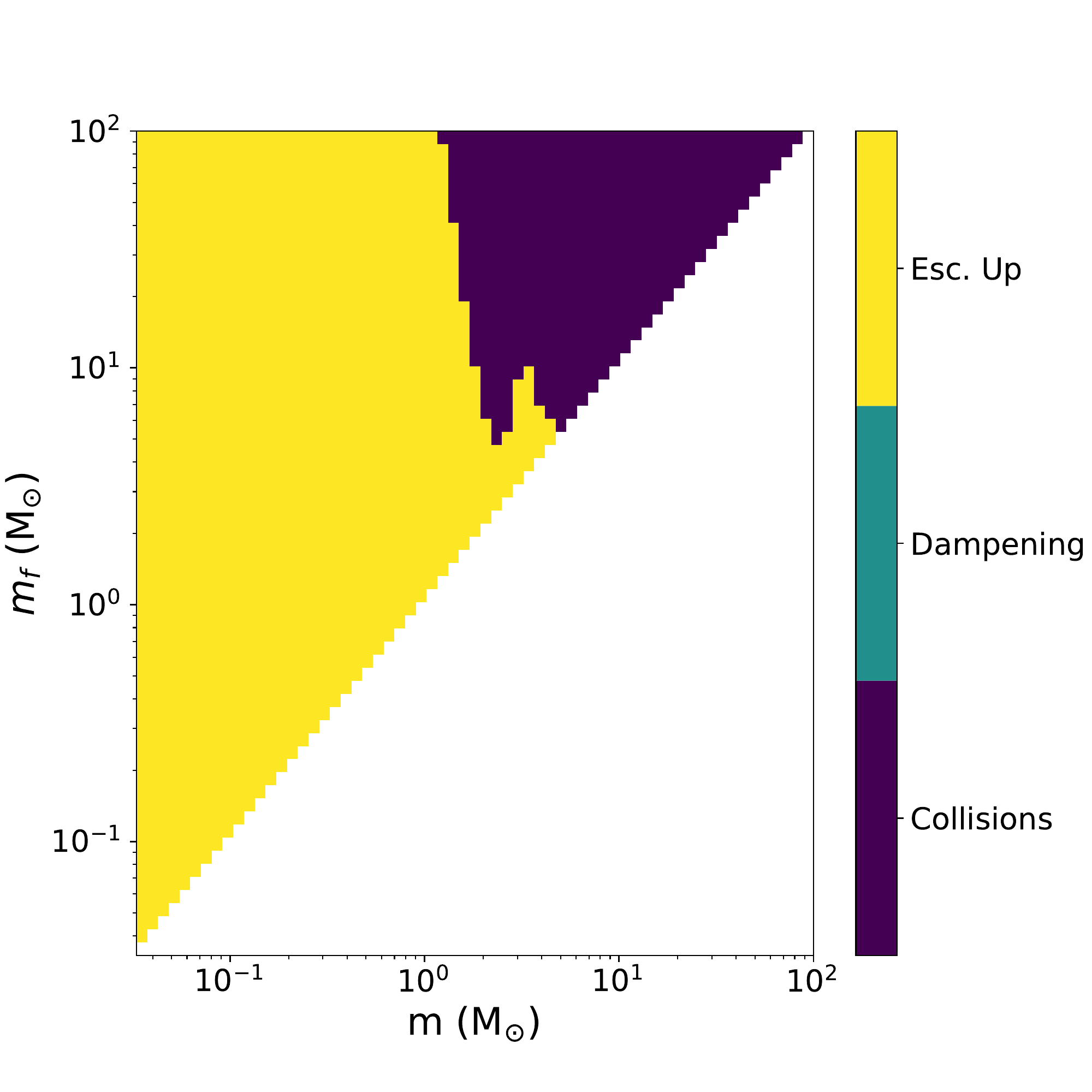}
\caption{\label{fig:sup_emax} Left: Maximum energy of proton CRs in units of GeV as a function of instantaneous mass, $m$, and final mass, $\mf$. Right: The dominant constraint on acceleration as a function of $(m, m_f)$, where ``Esc. Up" refers to upstream escape diffusion, ``Dampening" refers to wave dampening and ``Collisions" refers to interactions with H$_2$. We adopt the fiducial parameters: $B$ = 10 G, $\eta = 10^{-5}$ and f$_{acc}$ = 0.1.}
\end{figure*}

Figure \ref{fig:sup_emax} shows the maximum energy of CR protons as a function of protostellar mass and final mass and the dominant constraint on acceleration. We find for protostars with $m > 1 $ M$_{\odot}$, CR protons have maximum energies greater than 10 GeV. Only protostars with $M < 0.1$ M$_{\odot}$ have sub-GeV maximum energies. The maximum energy for solar and supersolar mass protostars is a constant E$_{\rm max} = 17$ GeV. This behavior changes at the transition between different acceleration constraints. CR acceleration in subsolar mass protostars is constrained by upstream diffusion. In this process, CRs are lost by diffusion upstream at the shock, thus inhibiting the maximum possible energy. At greater masses, the constraint is set by interactions with neutral gas near the shock. The collisional timescale becomes less than the time to accelerate CRs with $E > 17$ GeV.

The attenuated spectrum is only weakly affected by the cluster mass surface density, $\Sigma_{\rm cl}$ (Equation \ref{eq:sigcore} and \ref{eq:mdotnorm}). While a drop by a factor of 10 in $\Sigma_{\rm cl}$  produces a reduction by a similar factor in the unattenuated spectrum, the lower column results in a decline of a factor of only a few in the attenuated spectrum. It is also important to note that the unattenuated spectrum here is from the protostellar surface, while previous theoretical models have calibrated their CR spectrum from terrestrial or space-based measurements \citep{rab2017, rodgers-lee2017} and correct for geometric attenuation. However, it is difficult to correct for the effects of matter interactions. 

\subsubsection{Cosmic Ray Pressure}\label{pressec}

\begin{figure*}
\plottwo{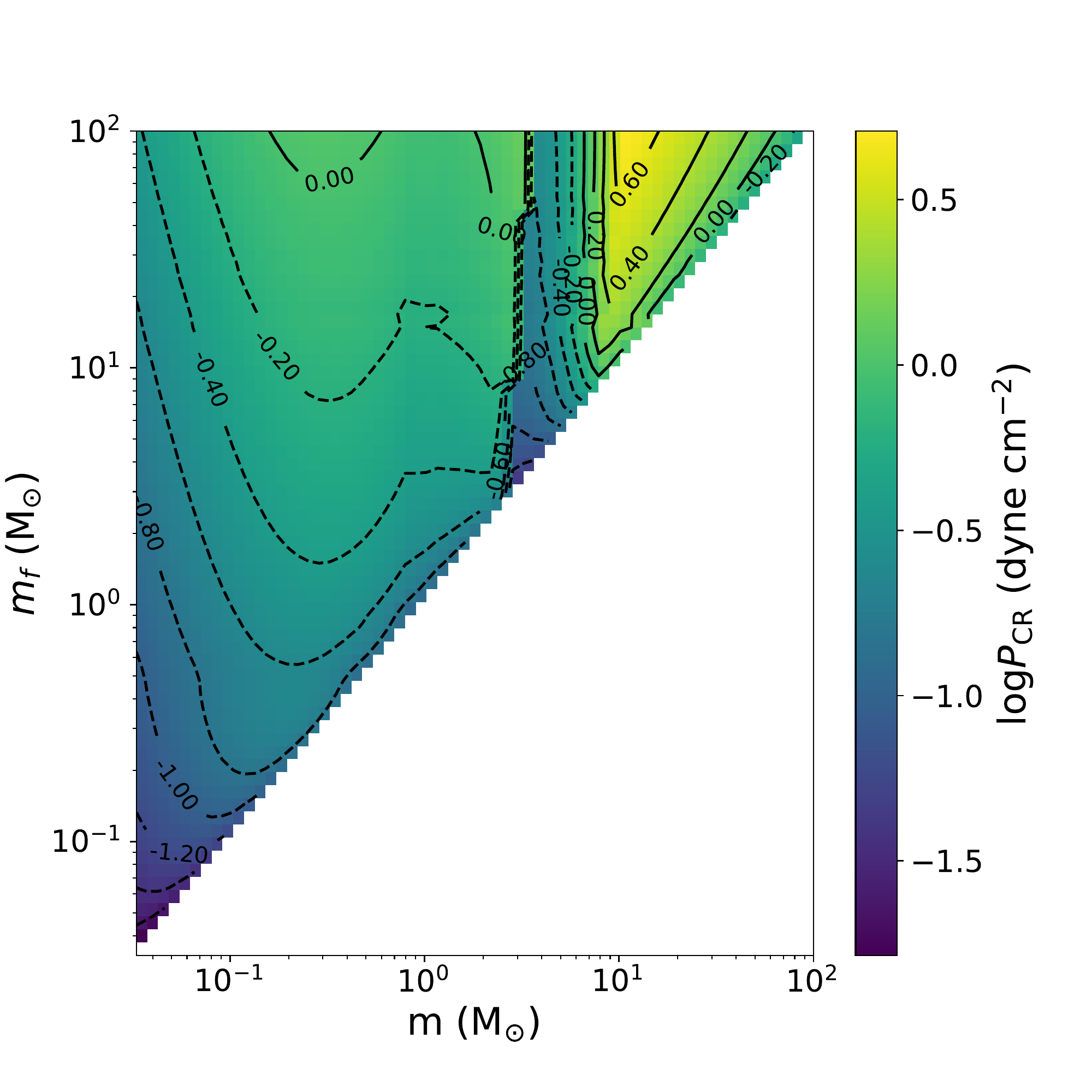}{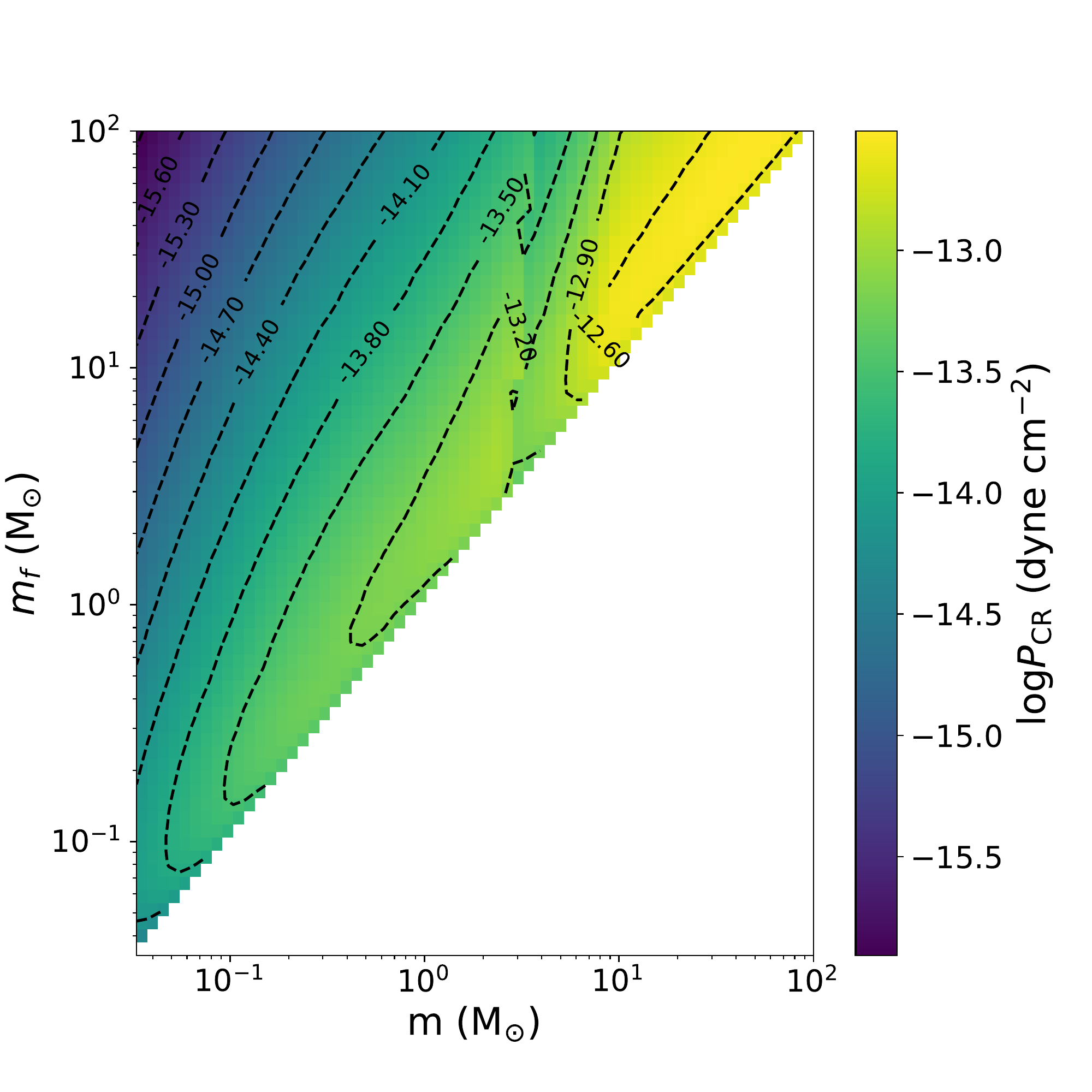}
\caption{\label{fig:pcr} Log cosmic ray pressure as a function of mass and final mass in units of dyne cm$^{-2}$. Left: Unattenuated cosmic ray pressure. Right: Attenuated cosmic ray pressure including matter interactions and geometric dilution. We set $\Sigma_{\rm cl} = 1.0$ g cm$^{-2}$.}
\end{figure*}

In order to properly model protostellar cores and to describe their dynamical state, various pressures must be taken into account. We calculate the CR pressure, $P_{\rm CR}$, from the energy flux spectrum:
\beq\label{eq:pcr}
P_{\rm CR} = \frac{4 \pi}{3} \int p(E) j(E) dE,
\eeq
where $p(E)$ is the relativistic momentum. 
Figure \ref{fig:pcr} shows the CR pressure across the parameter space of instantaneous mass, $m$, and final mass, $\mf$, assuming $\Sigma_{\rm cl} = 1.0$ g cm$^{-2}$. The unattenuated CR pressure is of order 1 dyne cm $^{-2}$ for most of the parameter space. There is a discrete change in the pressure at 3 M$_{\odot}$ caused by the similarly discrete radius change (itself brought on by a change in the internal structure of the protostar \cite{offner2011}). The pressure, in general, increases with mass. The maximum occurs when $m \approx 10$ M$_{\odot}$ and $\mf = 100$ M$_{\odot}$. The attenuated spectrum shows a significant decrease in the pressure: by 13 orders of magnitude. The gradient inverts and the highest CR pressures occur towards the $m = \mf$ boundary and towards the highest $(m,\mf)$. This is due to the change in the radius of the core. For a given final instantaneous mass, the core is physically smallest when the final mass is smallest. The discrete change at 3 M$_{\odot}$ is still apparent, although it is less significant. 

\begin{figure*}
\plottwo{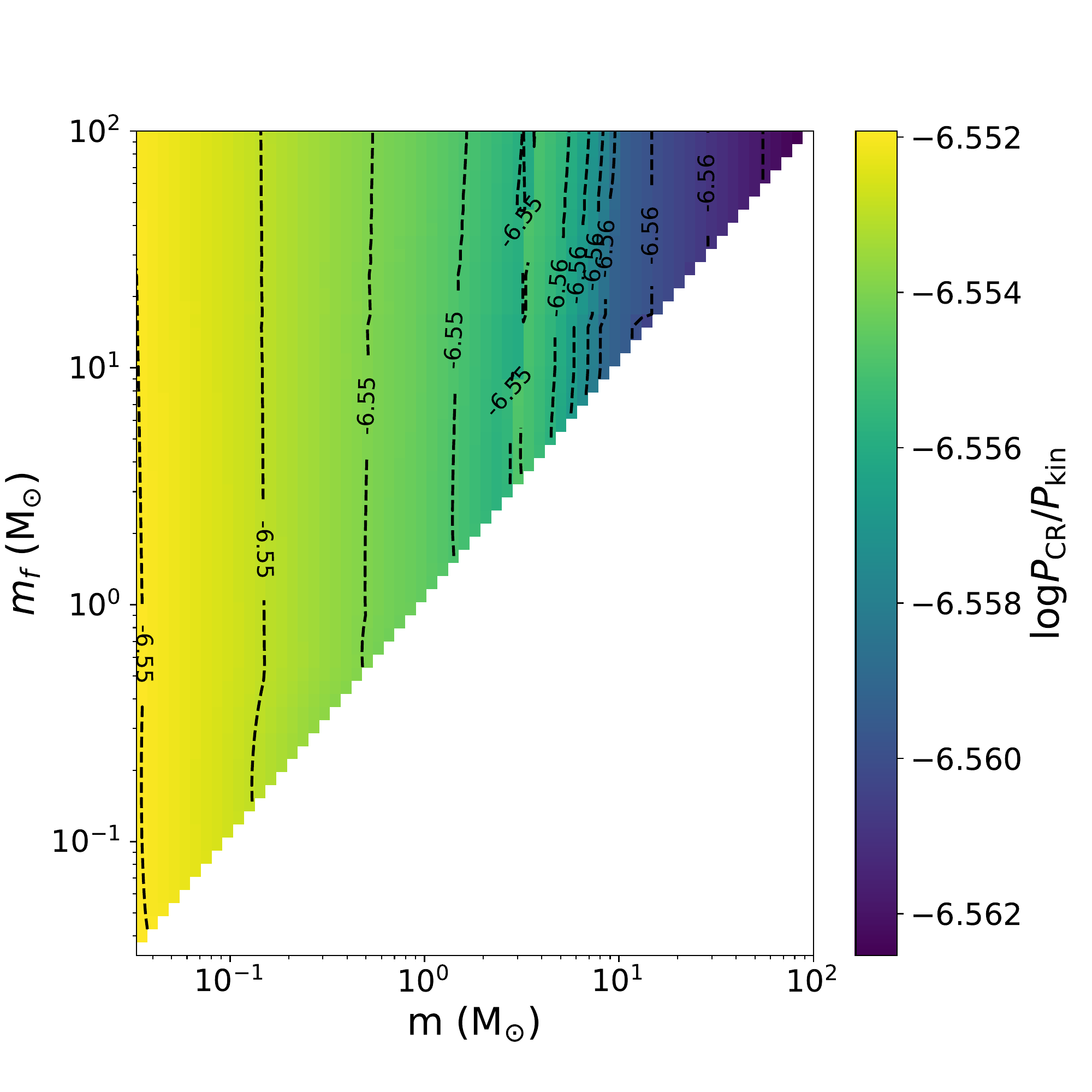}{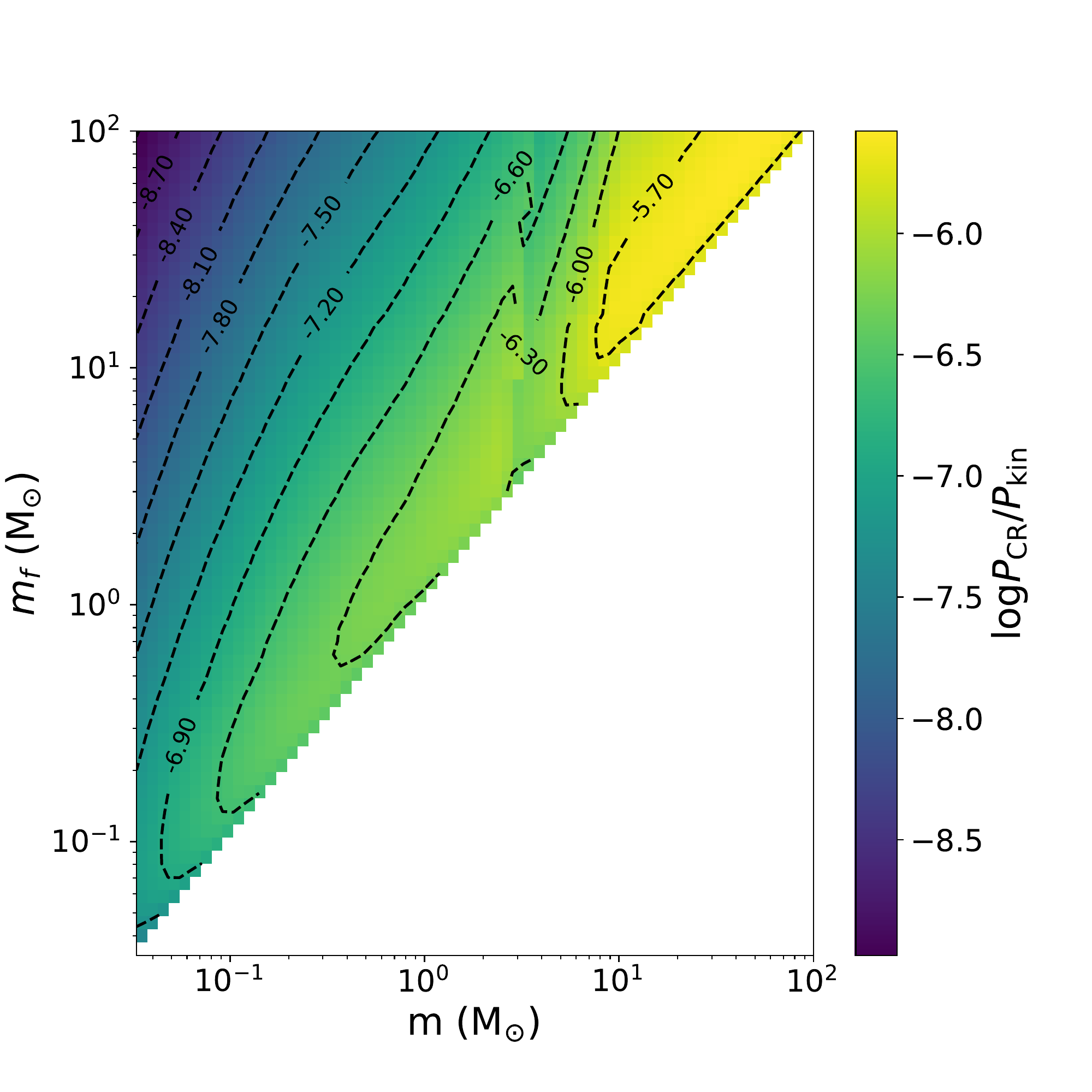}
\caption{\label{fig:prat} Log Ratio of CR pressure to kinetic pressure as a function of mass and final mass. Left: Unattenuated cosmic ray pressure, where the ram pressure is $P_{\rm kin} = \rho_{s} v_s^2$. Right: Attenuated cosmic ray pressure including matter interactions and geometric dilution, where the ram pressure is $P_{\rm kin} = \Phi_{\rm core}\Phi_s G \Sigma_{\rm cl}^2$. We set $\Sigma_{\rm cl} = 1.0$ g cm$^{-2}$.} 
\end{figure*}

The ratio of the CR pressure to the kinetic pressure is an important  test of the model. Figure \ref{fig:prat} shows the ratio $P_{\rm CR}/P_{\rm kin}$ across the $(m, \mf)$ parameter space.  We compare the unattenuated CR pressure to the ram pressure of the accreting matter, $P_{\rm kin} = \rho_s v_s^2$. Across the parameter space, $P_{\rm CR}$ is approximately a millionth of the kinetic pressure. Therefore, it is negligible compared to the accretion, as expected. The important kinetic pressure for the attenuated CR pressure is the surrounding molecular cloud turbulent pressure, $P_{\rm kin} = \Phi_{\rm core}\Phi_s G \Sigma_{\rm cl}^2$ with $\Phi_{\rm core} = 2$ and $\Phi_s = 0.8$ following \cite{mckee2003}. The maximum value of the ratio throughout the parameter space is only $P_{\rm CR}/P_{\rm kin} \approx 10^{-6}$. The CR pressure of CRs leaving the core is negligible to the dynamics of the surrounding molecular cloud as expected.


\subsection{Cosmic Ray Ionization Rates}\label{CRIRsec}

\subsubsection{Single Protostar}\label{CRIRsingle}
\begin{figure*}
\plottwo{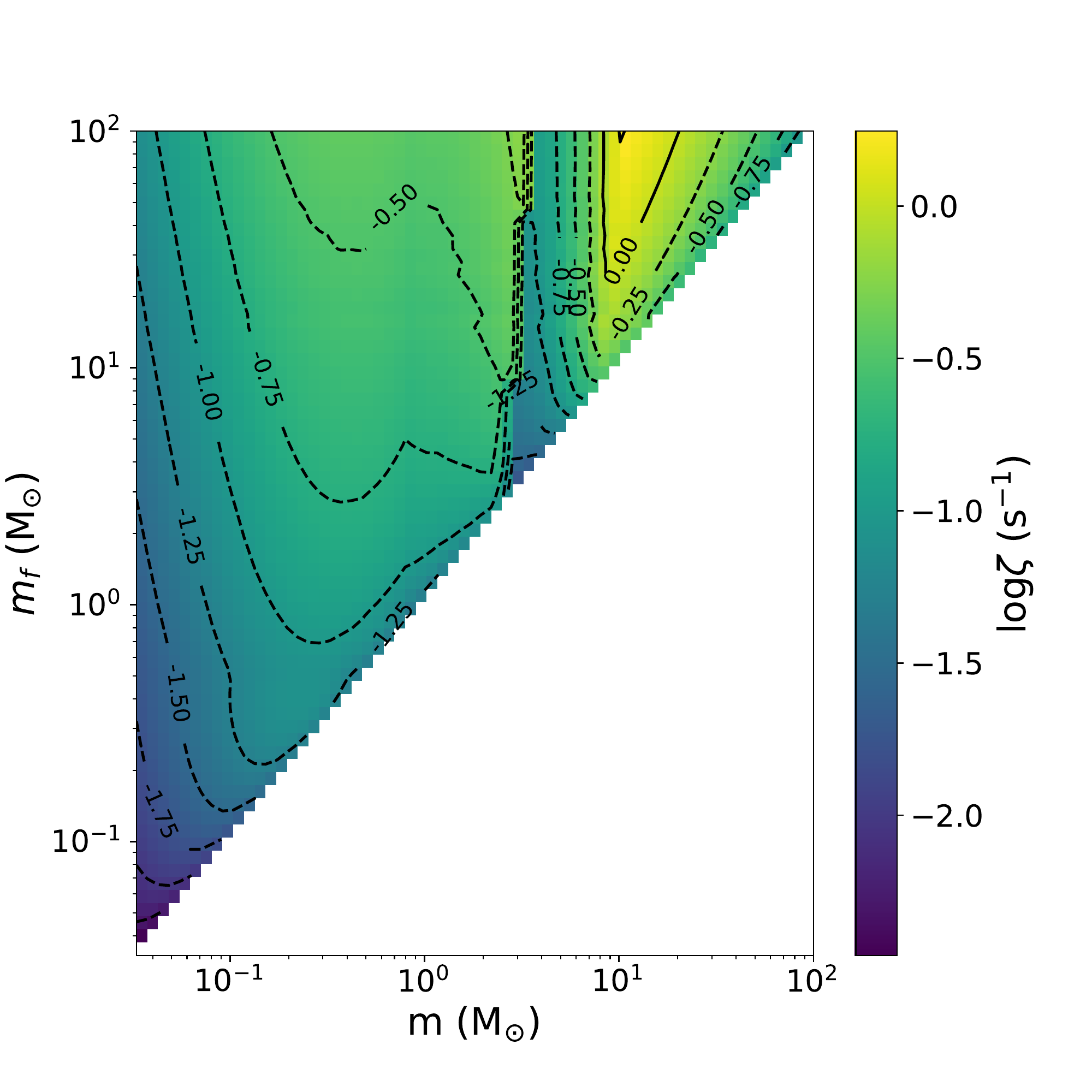}{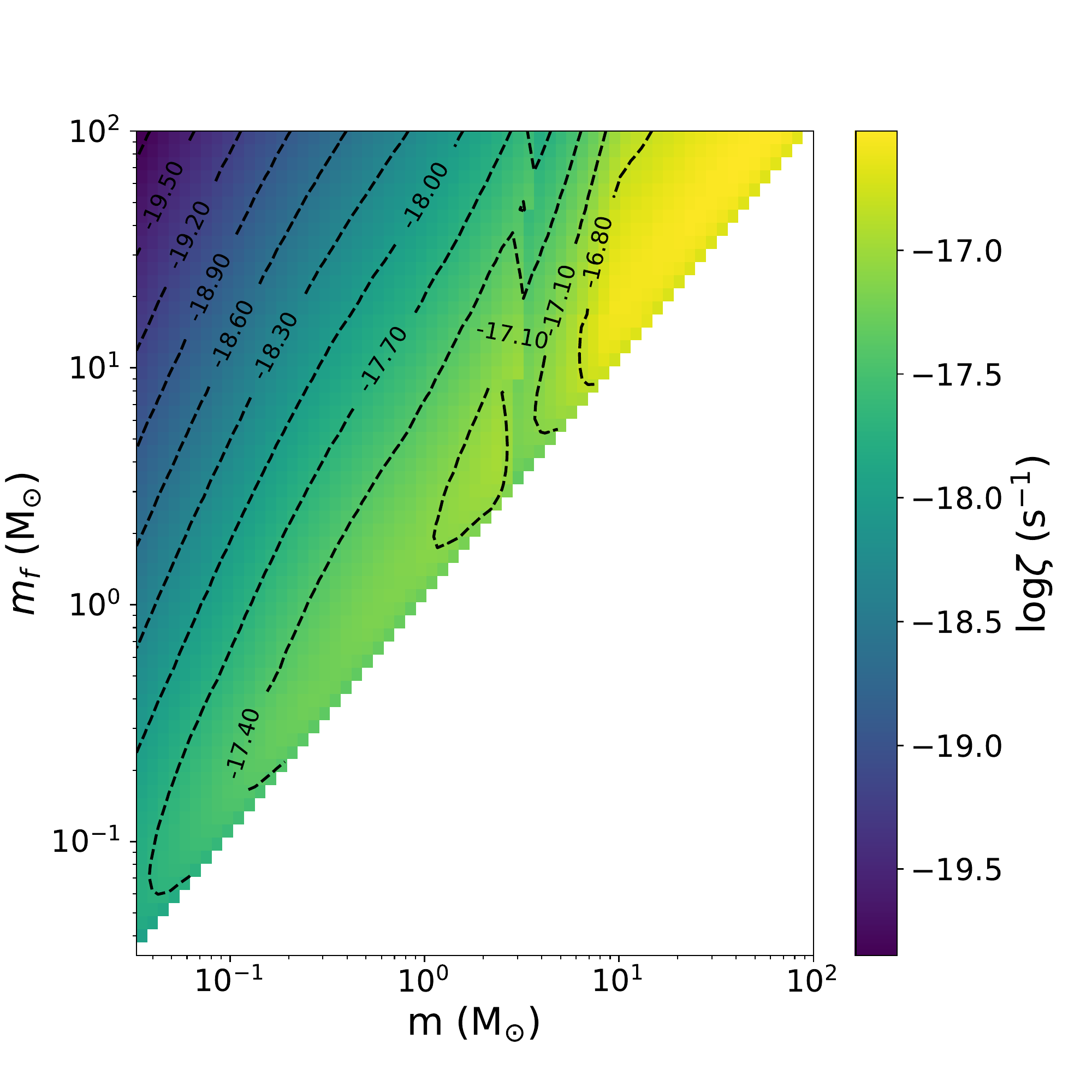}
\caption{\label{fig:zeta} Log cosmic ray ionization rate as a function of protostellar mass and final mass. Left: Cosmic ray ionization rate at the accretion shock. Right: Attenuated cosmic ray ionization rate at the edge of the protostellar core including matter interactions and geometric dilution. We set $\Sigma_{\rm cl} = 1.0$ g cm$^{-2}$.}
\end{figure*}

The CRIR is one of the key parameters of any astrochemical model, controlling the ionization fraction of gas with $A_V > $ a few, where the external FUV cannot penetrate. Figure \ref{fig:zeta} shows the CRIR, $\zeta$, as a function of $(m, \mf)$ for a single protostar. The same discrete jump at 3$_{\odot}$ appears due to the radius discontinuity discussed in \S\ref{pressec}. 

The unattenuated CRIR, near the protostellar surface, is incredibly high. Most of the parameter space exhibits $\zeta = 0.1 - 1$ s$^{-1}$. This value serves as an initial condition to scale the CRIR throughout the protostellar core. The attenuated CRIR, on the right side of Figure \ref{fig:zeta}, shows much more modest values. The attenuated CRIR at the surface of the core varies between $10^{-17} - 10^{-19}$ s$^{-1}$. The reduction is due in part to the radial dilution (decreasing the overall flux) and the collisional losses (moving 100 MeV - 1 GeV protons to lower energies ionize less efficiently). At the surface of the core,  the CRIR produced by an individual protostar becomes comparable to the attenuated external CRIR \citep{padovani2009}. Therefore, it is likely that in star-forming clouds, gas near embedded protostars may be equally affected by external and internal CR sources.

\begin{figure}
\plotone{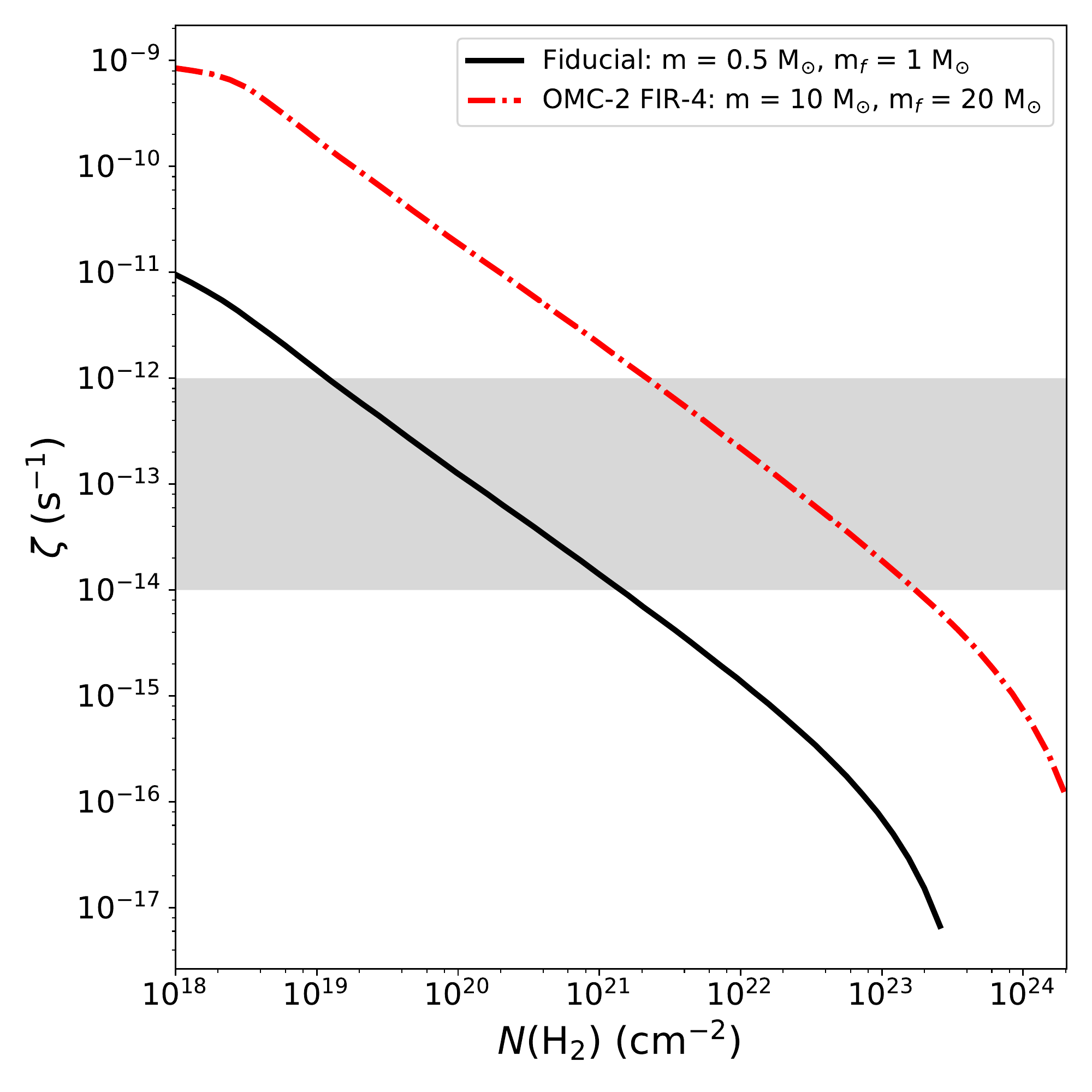}
\caption{\label{fig:zeta_vs_col} Cosmic ray ionization rate as a function of column density for a single protostar. The solid black line represents a protostar with instantaneous mass $m = 0.5$ M$_{\odot}$ and final mass $\mf = 1$ M$_{\odot}$ using the fiducial values in Table \ref{tab:params}. The dashed-dot red line represents a protostar with instantaneous mass $m = 10$ M$_{\odot}$ and final mass $\mf = 20$ M$_{\odot}$ with $\Sigma_{\rm cl} = 8$ g cm$^{-2}$. The grey box represents the order of magnitude range of $\zeta$ measured in OMC-2 FIR-4 by \cite{ceccarelli2014}.}
\end{figure}

There is a large difference between the unattenuated and the attenuated CRIR: 17 orders of magnitude. Figure \ref{fig:zeta_vs_col} shows the CRIR as a function of column density for a protostar with $m = 0.5$ M$_{\odot}$ and $\mf = 1.0$ M$_{\odot}$ as the solid black line. There is a near power-law behavior showing a 6 dex decrease in $\zeta$ with a 5 dex decrease in $N({\rm H_2})$. The column density is a proxy for the distance from the central protostar. As such, different molecules used to constrain $\zeta$ may suggest very different values of $\zeta$ depending on what radial surface they trace. 

Figures \ref{fig:spec}- \ref{fig:zeta_vs_col} assume a fixed value of $\Sigma_{\rm cl} = 1$ g cm$^{-3}$. $\Sigma_{\rm cl}$ has a linear relationship with the unattenuated CRIR. A decline of a factor of 10 in $\Sigma_{\rm cl}$ incurs a similar factor of 10 decrease in the unattenuated $\zeta$ and P$_{\rm CR}$ due to the decrease in the accretion rate through $\dot{m}_{\rm TC}$. However, there is a much weaker dependence of $\Sigma_{\rm cl}$ on the attenuated CRIR and CR pressure. The core radius depends on $\Sigma_{\rm cl}^{\frac{1}{2}}$ and the core molecular column density $N({\rm H_2})$ scales linearly with $\Sigma_{\rm cl}$. Together, these factors result in only a factor of a few decrease in $\zeta$ and $P_{\rm CR}$ with an order of magnitude decrease in $\Sigma_{\rm cl}$. We present $\zeta(N)$ for the whole $(m, \mf$) space and for different values of $\Sigma_{\rm cl}$ in an interactive online tool \footnote{\url{http://protostarcrs.brandt-gaches.space}\label{foot:site}}.

\subsubsection{Protostellar Cluster Cosmic Ray Ionization Rate}\label{ClusterCRIR}

We have so far presented results for individual protostars within their natal core. However, molecular clouds form many stars simultaneously. We have shown in \S\ref{CRIRsingle} that at the protostellar core surface, the CRIR can be on par with the attenuated external CRIR. This suggests it is important to consider both CRIR components in order to understand cloud chemistry in forming clusters.

\begin{figure*}
\plotone{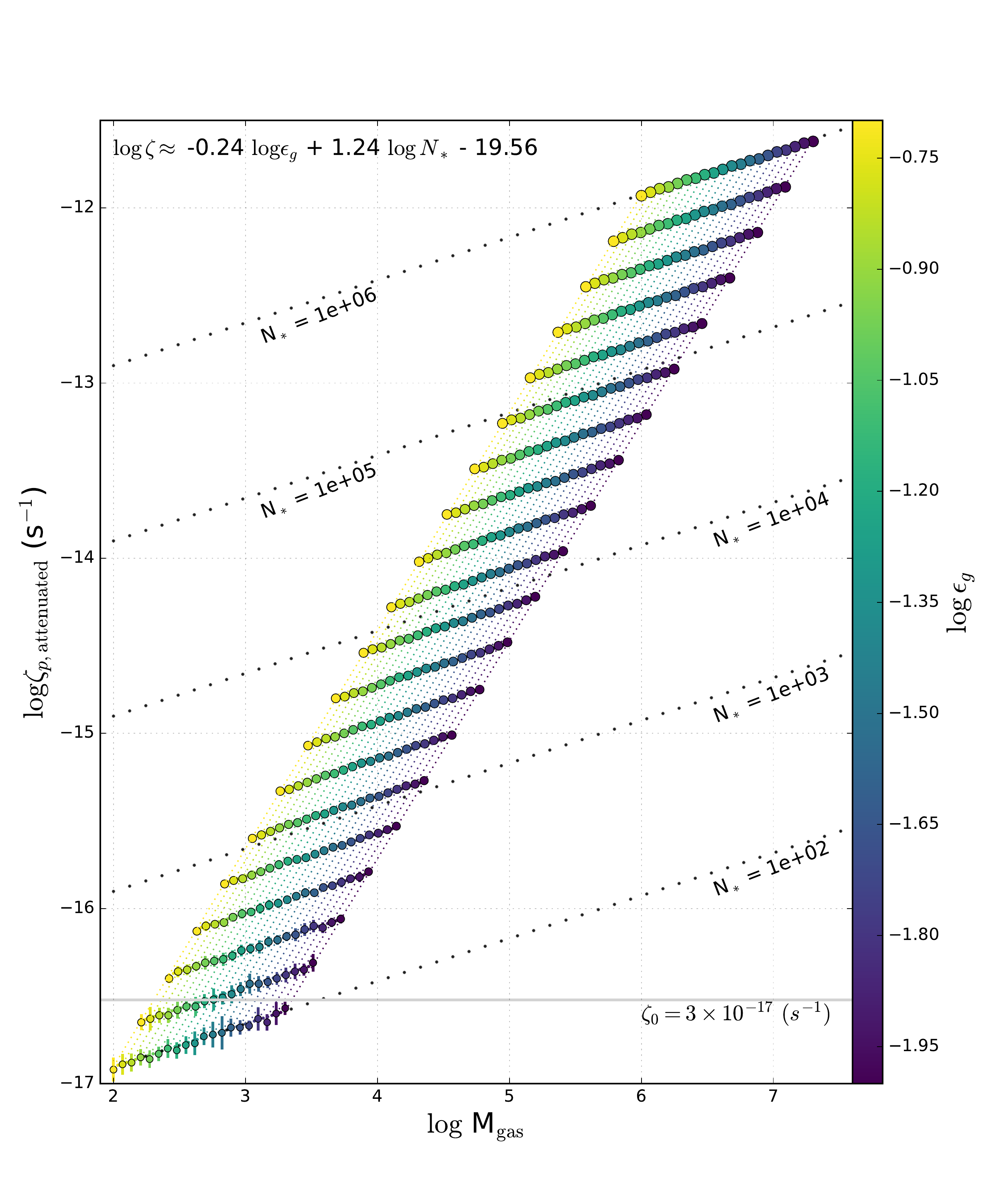}
\caption{\label{fig:clusterCRIR} Attenuated cosmic ray ionization rate as a function of number of protostars in the cluster, star formation efficiency and gas mass. Error bars indicate the $\pm 1 \sigma$ spread. Point size and color indicate $N_*$ and $\epsilon_g$, respectively. The gray horizontal line indicates the fiducial value $\zeta_0 = 3 \times 10^{-17}$ (s$^{-1}$). Dotted black lines show lines of constant $N_*$. A two-dimensional fit of $\log \zeta (N_*, \epsilon_g)$ is annotated in the top left corner.}
\end{figure*}

Figure \ref{fig:clusterCRIR} shows the attenuated CRIR due to all the embedded protostars in a cluster. The size of the points indicates the number of protostars and the color of the points indicates the assumed star formation efficiency, which impacts the result through $\Sigma_{\rm cl}$. Figure \ref{fig:clusterCRIR} represents 400 mock clusters covering a large range of $(N_*, \epsilon_g$). The error bars indicate the 1$\sigma$ spread due to sampling the bi-variate PMF. For $N_* > 500$, the error bars are smaller than the data points due to more complete sampling of the bi-variate PMF. 

The two parameters, $\epsilon_g$ and $N_*$, produce opposite trends in the CRIR. A reduction in $\epsilon_g$ leads to a higher $\Sigma$, thus causing a greater CRIR. However, this effect is sublinear - a dex change in $\epsilon_g$ leads to less than a dex change in the cluster CRIR. The CRIR depends more strongly on N$_*$ than $\epsilon_g$. Increasing N$_*$ leads to a slightly super-linear increase in $\zeta$ due to the inclusion of more high-mass protostars.

We fit $\zeta(N_*, \epsilon_g)$ with a two-dimensional linear function in log space, which represents the model results well:
\beq\label{eq:cluster_res}
\log \zeta = -0.24 \log \epsilon_g + 1.24 \log N_* - 19.56.
\eeq
We plot this function on Figure \ref{fig:clusterCRIR} as the dashed lines, and we add lines of constant N$_*$ for reference. For clusters with N$_* > $ few hundred protostars, the CRIR due to embedded protostars is {\it greater} than the typically assumed fiducial rate of $\zeta_0 = 3 \times 10^{-17}$ s$^{-1}$, which is shown as the gray solid line.


\section{Discussion}\label{disc}

\subsection{Variations of Physical Parameters}

There are 3 key unknowns in our model: the protostellar magnetic field, $B$, the accretion flow filling fraction, $f$ and the shock efficiency parameter, $\eta$.  We discuss the uncertainties and impact of each below.

\subsubsection{Magnetic Field Strength}

The magnetic field strength at the protostellar surface is not well constrained. Typically, it is thought to range between a few Gauss and 1 kG. In our fiducial model, we assume $B$ = 10G. However, this is on the smaller end of the possible range. The magnetic field plays a dominant role in setting the maximum energy due to wave dampening. Wave dampening is very sensitive to the magnetic field. A small increase in the magnetic field leads to significantly more dampening of self-produced Alfv\'{e}n waves by the CRs, described in detail in Appendix \ref{CRphysics}. The dampening criterion in Equation \ref{eq:edamp} depends on $B^{-4}$; a factor of hundred difference in magnetic field strength yields a substantial change in this criterion. 

\begin{figure*}
\plottwo{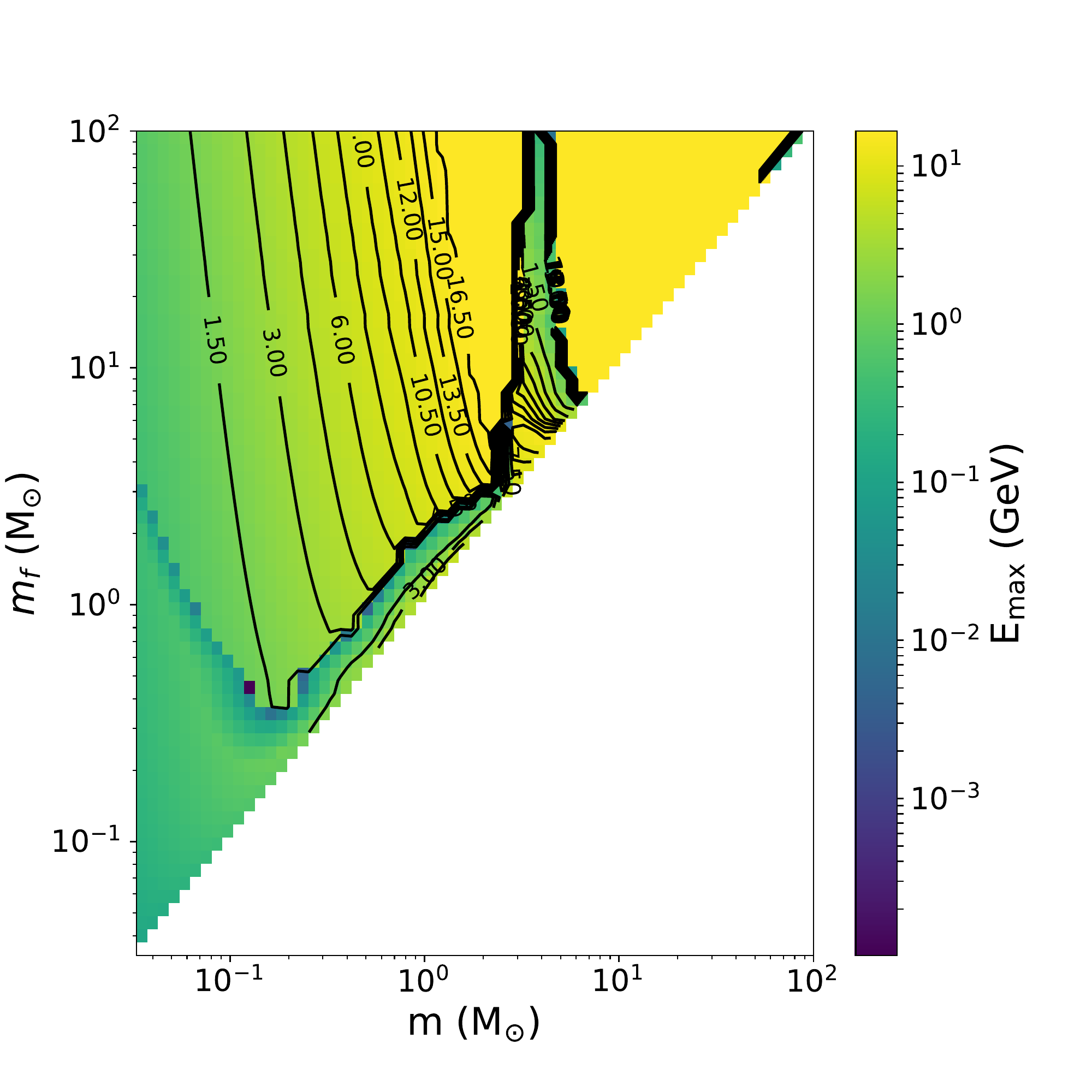}{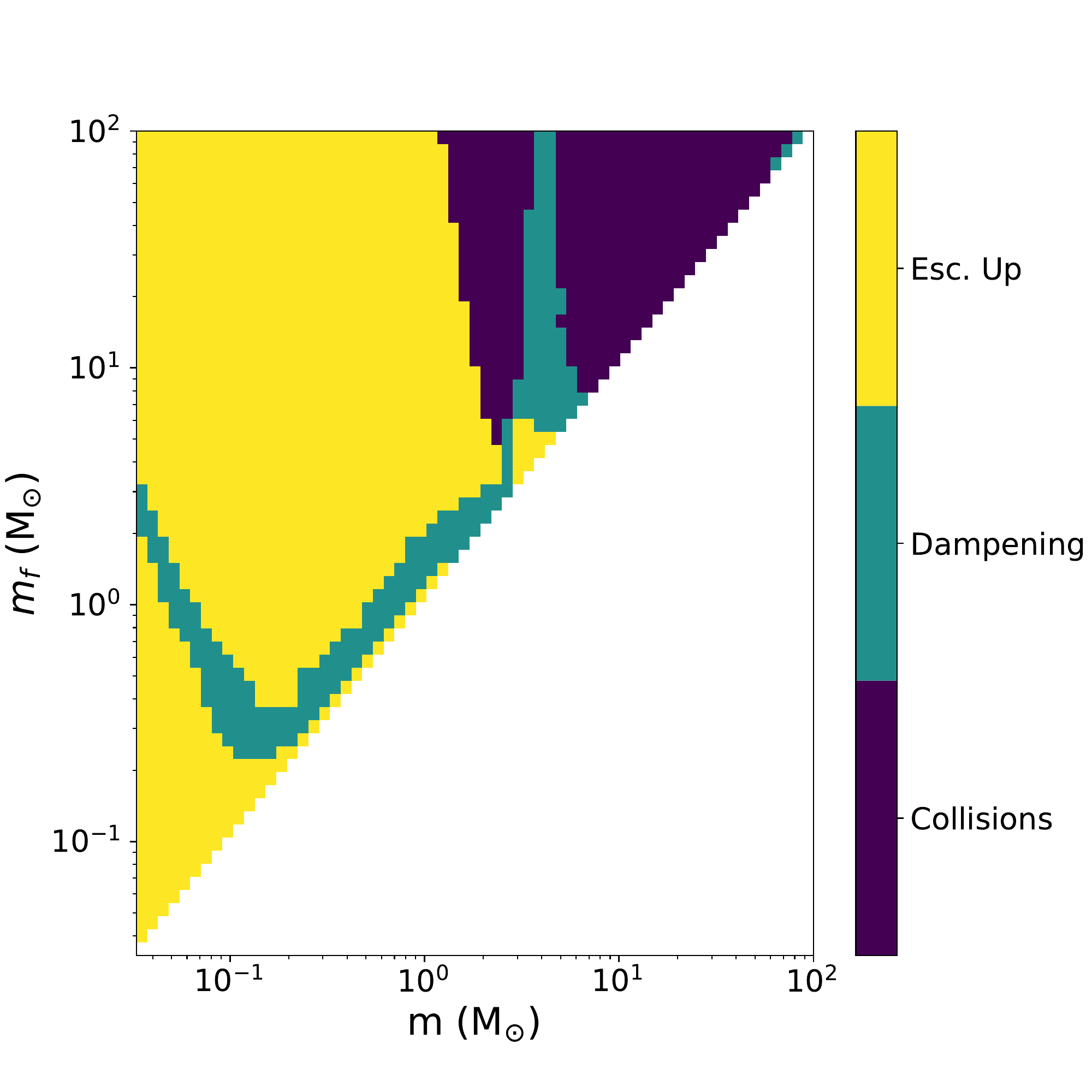}
\caption{\label{fig:sup_emax_1kG} Same as Figure \ref{fig:sup_emax} but with B = 1 kG.}
\end{figure*}

We recalculated the CR spectrum and CRIR for the $(m, \mf)$ parameter space with B = 1 kG. Figure \ref{fig:sup_emax_1kG} shows the maximum energy and acceleration constraints as a function of $(m, \mf)$. We find that there are swaths of the parameter space where the shock density, temperature and velocity are such that wave dampening becomes the dominant constraint in acceleration. Figure \ref{fig:sup_emax_1kG} shows that in these regions E$_{\rm max}$ is reduced to 50-100 MeV significantly below the GeV energy scale in our fiducial model. This has relatively little effect on the unattenuated CRIR. However, for high column densities, the collisional losses are sufficient to significantly reduce the high CR flux. Therefore, there are regions within the $(m, \mf)$ parameter space where the attenuated CRIR will be negligible due to wave dampening. These regions only account for a few percent of the parameter space ,i.e. mainly low-mass protostars, so that our cluster results are largely independent of the assumed magnetic field strength.


\subsubsection{Accretion Flow Filling Fraction}

\begin{figure*}
\plottwo{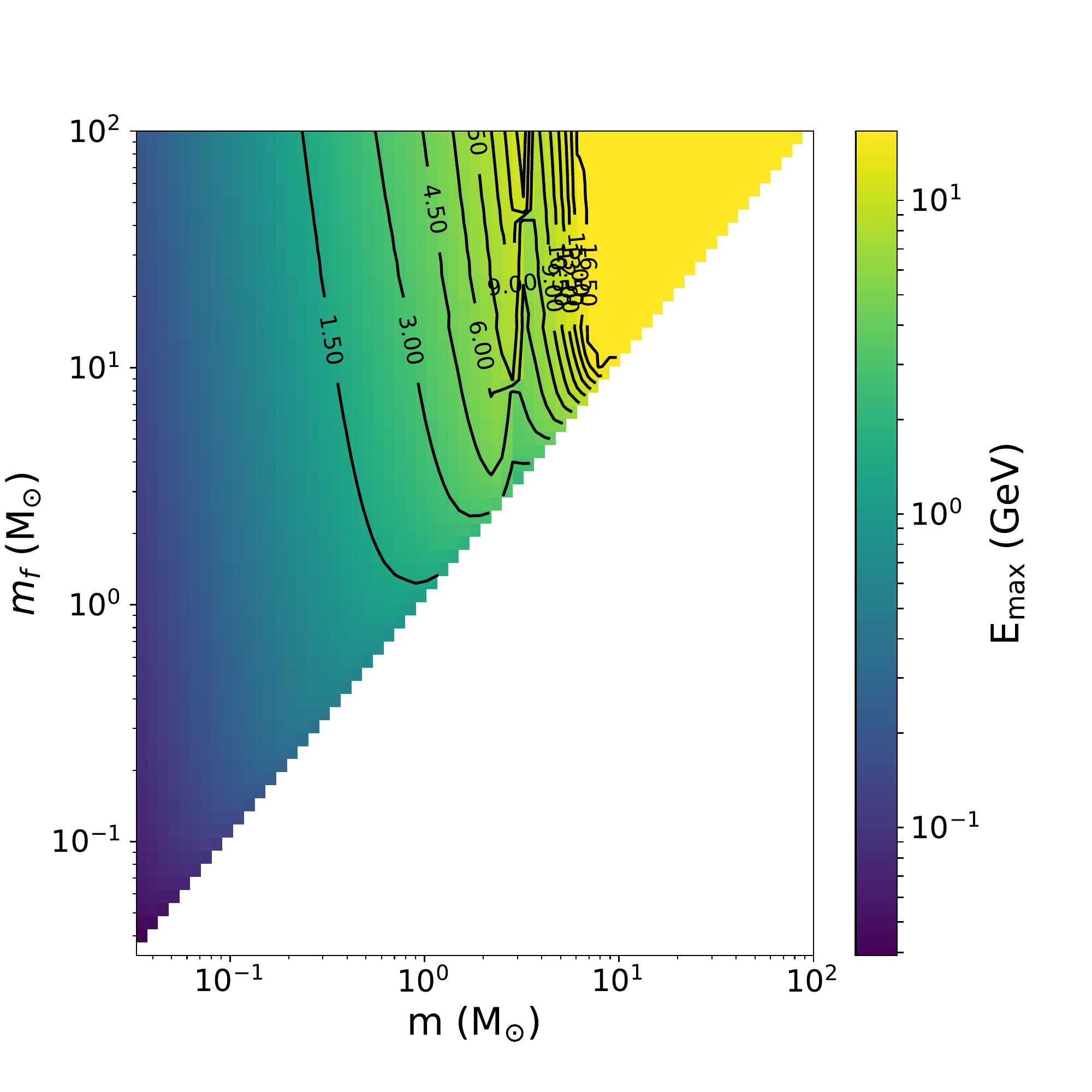}{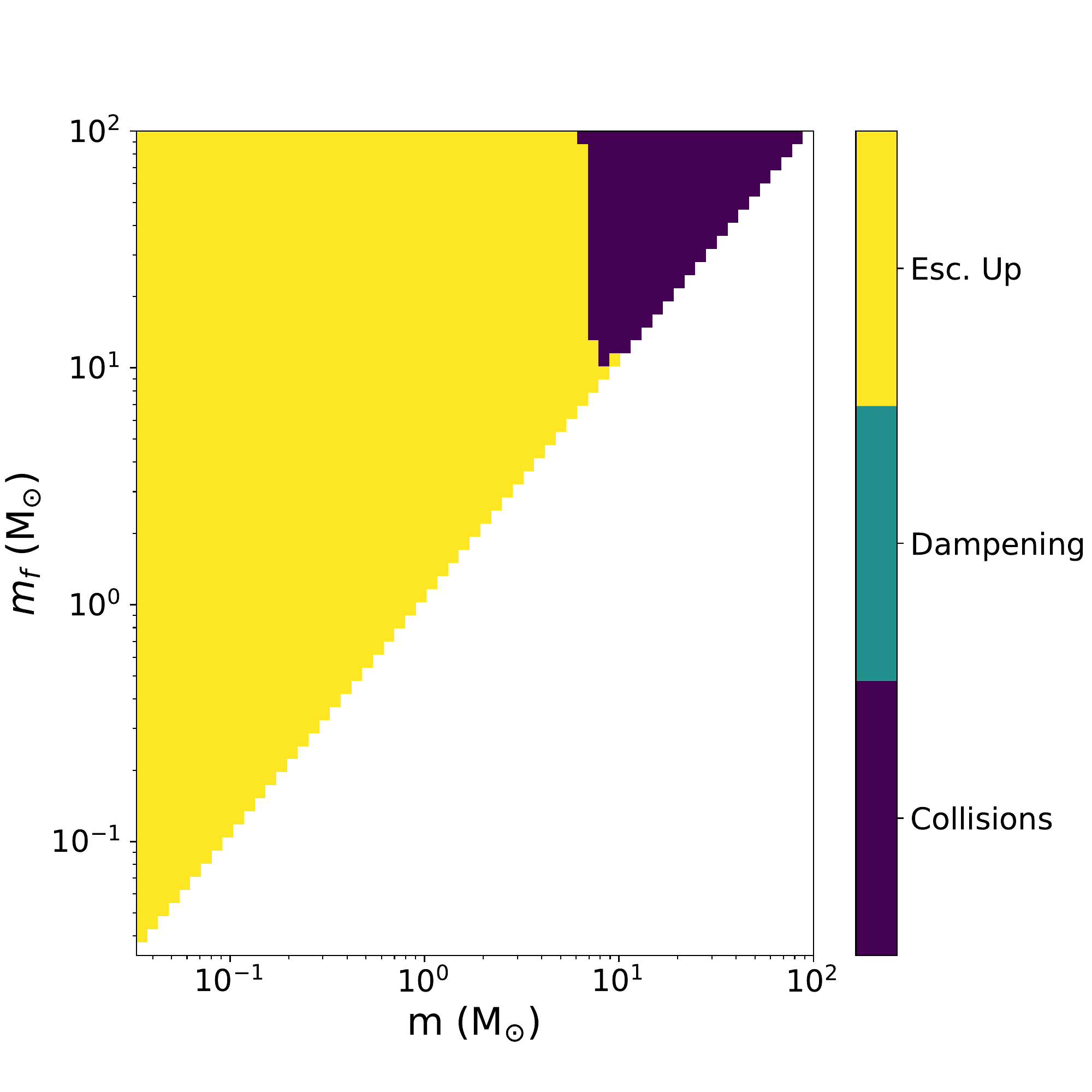}
\caption{\label{fig:sup_emax_facc} Same as Figure \ref{fig:sup_emax} but with $f_{\rm acc} - 0.9$.}
\end{figure*}

The accretion flow filling fraction, $f$, directly influences the shock density as $n \propto f^{-1}$. Our fiducial model assumes $f = 0.1$. However, Class 0 sources, which likely have higher accretion rates, may undergo more spherical accretion. We investigate the effect of increasing the shock filling fraction to $f = 0.9$. Figure \ref{fig:sup_emax_facc} shows the maximum energy and accleration constraint mechanisms for $f = 0.9$. The behavior is very similar to the fiducial values in Figure \ref{fig:sup_emax}, although the region where the acceleration is constrained by matter interactions is smaller and pushed towards higher masses. We find that a factor of 9 increase in $f$ leads to a factor of 3-4 decrease in $E_{\rm max}$ for protostars with masses below 3 M$_{\odot}$. The maximum energy for the rest of the parameter space remains above 10 GeV. For this higher filling fraction, we find a factor of 9 decrease in the unattenuated and attenuated CRIR due to a change in the normalization of $f(p)$ ( see Equation \ref{eq:fp}).

While variation in the filling fraction leads to a respectively linear change in the CRIR, in a cluster environment higher $f$ values for young sources may cancel with lower values exhibited by older sources, whose accretion has declined. 

\subsubsection{Shock Efficiency Parameter}

\begin{figure*}
\plottwo{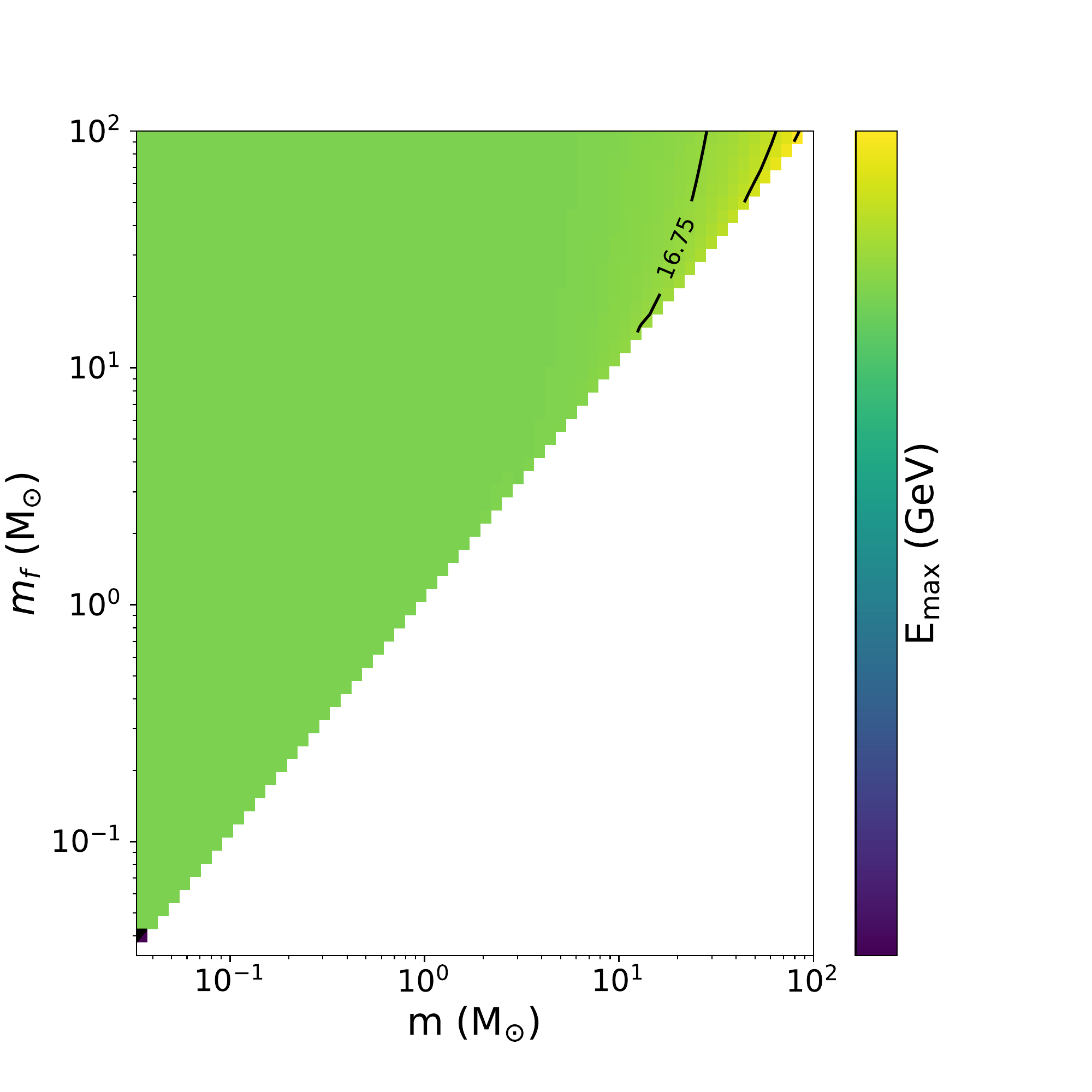}{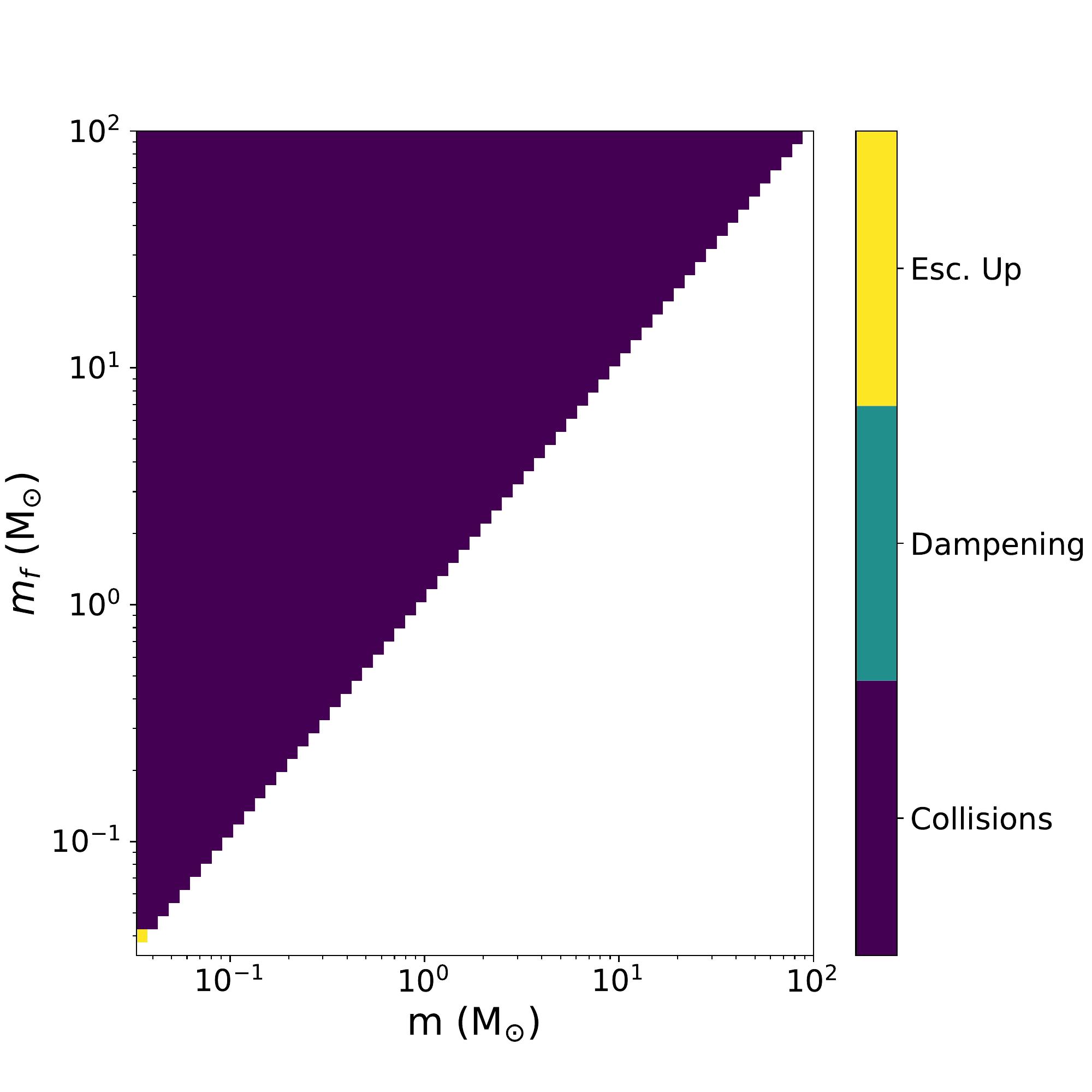}
\caption{\label{fig:sup_emax_eta} Same as Figure \ref{fig:sup_emax} but with $\eta = 10^{-3}$.}
\end{figure*}

The shock efficiency parameter, $\eta$, which describes the fraction of thermalized protons that are accelerated to relativistic speeds, is also  poorly constrained. We assume $\eta = 10^{-5}$ for the fiducial model following \cite{padovani2016}. The normalized CR pressure, $\tilde{P}_{\rm CR} = \frac{P_{\rm CR}}{\rho_s v_s^2}$, depends linearly on $\eta$ (see Appendix \ref{CRphysics} for details). Stronger CR pressure, in relation to the shock ram pressure, decreases the effectiveness of wave dampening. As such, the maximum energy is constrained mainly by interactions with neutral gas. Figure \ref{fig:sup_emax_eta} shows the maximum energy and acceleration constraints for $\eta = 10^{-3}$. We find that the maximum energy for the whole parameter space is approximately 17 GeV. A 2 dex increase in $\eta$ leads to a 2 dex increase in the unattenuated and attenuated CRIR,  i.e., $\zeta$ depends linearly on $\eta$. Consequently, uncertainties in $\eta$ lead to large uncertainties in the expected CRIR. However, evidence for CRs with energies greater than 10 GeV would be an indication of a higher $\eta$.

\subsubsection{Transport Parameter}\label{sec:transport}

\begin{figure}
\plotone{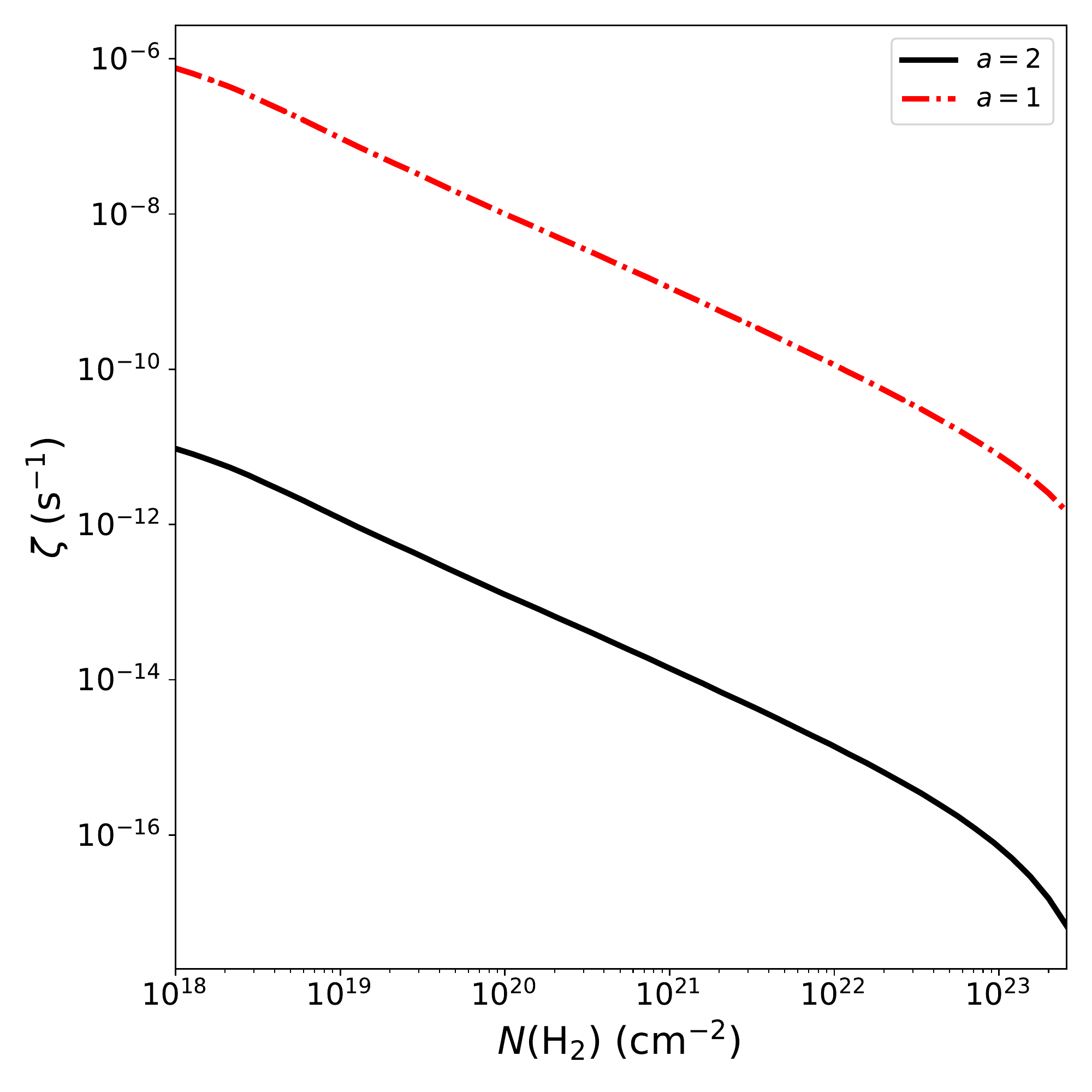}
\caption{\label{fig:zeta_transport} Cosmic ray ionization rate as a function of column density from a single protostar. The solid black line indicates free-streaming transport with $a=2$. The dashed-dotted red line shows the case for diffusive transport. The parameters assumed for the plot are $m = 0.5$ M$_{\odot}$, $\mf = 1.0$ M$_{\odot}$ and $\Sigma_{\rm cl} = 1$ g cm$^{-2}$.}
\end{figure}

How CRs are transported through a protostellar core from its central protostar has not been modeled in detail. Transport of CRs is determined by factors relating to the gas density, magnetic field configuration, diffusion coefficients and cosmic ray energy \citep{padovani2013, rodgers-lee2017}. Figure \ref{fig:zeta_transport} shows a comparison between the two limiting cases of transport through the core for a subsolar Class 0 protostar. The CRIR is five orders of magnitude higher in the diffusive regime than the free streaming. At the edge of the core, the CRIR is $\zeta = 10^{-11}$ s$^{-1}$. Balancing cosmic ray heating, $\Gamma_{\rm cr}$, with atomic and molecular line cooling, given by \cite{goldsmith2001}, predicts temperatures of $T > 10^3$ K for densities of $n=10^3$ cm$^{-3}$ and a CRIR $\zeta = 10^{-11}$ s$^{-1}$. Such temperatures at the core edge are inconsistent with observations. However, $\zeta = 10^{-17}$ s$^{-1}$, the case of free-streaming, produces temperatures of $T \approx 10$ K. Observations of molecular ions can measure the CRIR in the outer regions of cores, constraining the transport mechanism. We discuss this in \S\ref{sec:obs_comp}

The case of transport in protostellar disks is much more complicated. The transport through the disk strongly depends on assumptions about the magnetic field morphology \citep{padovani2018}. In contrast, protostellar core magnetic fields are thought to exhibit an hour glass morphology \citep{machida2007, crutcher2012}. Such a morphology will allow free streaming, although not fully isotropically.

\subsection{Comparison with Observations}\label{sec:obs_comp}

Directly measuring the CR flux from embedded protostars is not possible. However, the CRIR can be constrained through modeling the radio and sub-millimeter emission from molecular ions.
There have been several recent observations towards embedded protostars, which have attempted to constrain the CRIR.

\cite{ceccarelli2014} measured HCO$^+$, H$^{13}$CO$^+$ and N$_2$H$^+$ emission towards OMC-2 FIR-4. OMC-2 FIR-4 is a protocluster within the Orion Molecular Cloud (OMC) at a distance of 420 pc which contains a few low- and intermediate-mass protostars, a total mass of 30 M$_{\odot}$ and luminosity of 10$^3$ L$_{\odot}$ \citep{kim2008, crimier2009, lopez-sepulcre2013}. They  modeled the chemistry using a two zone model: a warm inner region and a cold envelope. The inner region, at a radius of 1,600 AU, is well fit by a CRIR of $\zeta = 6 \times 10^{-12}$ s$^{-1}$, while the outer envelope, at a distance of 3,700 AU, has a CRIR of $\zeta = 4 \times 10^{-14}$ s$^{-1}$. They use a power-law CR flux spectrum, f(E) $\propto E^p$, with $p$ between -4 and -2.5. The central compact source in OMC-2 FIR-4 is thought to be an early stage Class 0 protostar, with a mass around 10 M$_{\odot}$ \citep{lopez-sepulcre2013, furlan2014}. To compare with their results, we assume that this source dominates the CR flux and bolometric luminosity. Figure \ref{fig:zeta} shows that for a 10 M$_{\odot}$ protostar, the CRIR is fairly insensitive to the final mass. Figure \ref{fig:zeta_vs_col} shows the inferred CRIR with a protostar of $(m, \mf) = (10, 20)$ M$_{\odot}$ and $\Sigma_{\rm cl} \approx 8.0$ g cm$^{-2}$ following the column density measurements of \cite{lopez-sepulcre2013}. We show $\zeta \approx  10^{-12} - 10^{-14}$ between column densities of $2\times10^{21} - 2\times10^{23}$ cm$^{-2}$. Therefore, our model is consistent with the enhanced CRIR measured towards OMC-2 FIR-4 from CRs accelerated by the central protostar's accretion shock. Under these assumptions,  $\zeta > 10^{-14}$ at column densities $N({\rm H_2}) < 3\times10^{23}$ cm$^{-2}$. The elevated CRIR in OMC-2 FIR-4 has been similarly inferred from  HC$_3$N and HC$_5$N \citep{fontani2017} and c-C$_3$H$_2$ \citep{favre2018}. The observed CRIR towards OMC-2 FIR-4 is consistent with the free streaming approximation in \S\ref{CRIR_model}.

\cite{favre2017} expanded the work of \cite{ceccarelli2014} with a survey of Class 0 protostars spanning low to intermediate masses. They use high J transitions of HCO$^+$ and N$_2$H$^+$ to measure the ratio HCO$^+$/N$_2$H$^+$ to infer the CRIR. They assume a fixed temperature of 40 K and density of $2.5 \times 10^5$ cm$^{-3}$. They could not confirm a systematically higher CRIR in embedded Class 0 sources due to large errors in converting the molecular emission to an abundance. For many sources, the full Spectral Line Energy Distribution (SLED) of HCO$^+$ and N$_2$H$^+$ are not observed, and some sources are not detected in N$_2$H$^+$. However, \cite{favre2017} show that the ratio does not depend strongly on the luminosity of the protostar. In our results, we find that the parameter space for protostars between 0.1 M$_{\odot}$ and 3 M$_{\odot}$ exhibits a relatively flat $\zeta$ dependence. One caveat is that not all sources have all molecular lines detected so the emission may trace different column density surfaces. We show in Figure \ref{fig:zeta_vs_col} that this could result in orders of magnitude difference in $\zeta$. This makes it difficult to constrain the absolute value of $\zeta$ without constraining the radial surfaces and temperatures, as done in \cite{ceccarelli2014}.

\cite{cleeves2015} measured the total ionization rate towards TW Hya, which is an evolved Class II protostar. They found $\zeta_{\rm CR} <$ 10$^{-19}$ s$^{-1}$, which  is discrepant with our results. However, TW Hya has $\dot{m} \approx 10^{-9}$ M$_{\odot}$ yr$^{-1}$ \citep{ingleby2013}. Our results focus on Class 0 and Class I protostars, which are still accreting from their envelope.  Therefore, we would not expect CR acceleration to be efficient in this system. 

\section{Summary}\label{summary}

We present self-consistently derived CR spectra and CRIRs for protostars and protoclusters from accretion shocks at the protostellar surfaces. We combine a CR model \citep{padovani2016} with analytic accretion history models. We find that protostars are efficient accelerators of protons from energies between keV to GeV scales. The energy losses due to diffusion escape and collisional losses inhibit acceleration of CRs to TeV scales, indicating that gamma radiation would not be present. Furthermore, the CR flux spectrum is consistent with an ideal supersonic, super-Alfv\'{e}nic shock with $j(E) \propto E^{-2}$. Collisional losses due to envelope gas interactions and geometric dilution substantially decrease the CR flux at the edge of the envelope such that the spectrum at lower energies flattens. 

We quantify the CR pressure and the importance of this pressure to the kinetic pressure and find that the CR pressure is minimal, confirming that it need not be included in a virial analysis of cores.

We present the CRIR for protostars  for a broad range of instantaneous and final protostellar masses. Protostellar accretion shocks are efficient accelerators of CRs, producing $\zeta > 10^{-12}$ s$^{-1}$ in the inner region of their envelopes and disks. Towards the edge of the envelope, $\zeta$ drops to 10$^{-17}$ s$^{-1}$. However, within the natal molecular cloud, this rate is still greater than that due to external CR sources if collisional losses are accounted for \citep{padovani2009}. We present the results from this paper over an extended parameter space in an online interactive tool (See Footnote 1). {\it We conclude individual protostars may  dominate the high extinction gas ionization in their natal cloud. }

We calculate $\zeta$ for protoclusters as a function of the number of constituent protostars, N$_*$, and star formation efficiency, $\epsilon_g$. We find that protoclusters with N$_* \gtrsim$ a few hundred exhibit $\zeta$ greater than the often assumed value of $\zeta_0 = 3 \times 10^{-17}$. Large protoclusters, such as those within the OMC, will accelerate CRs and provide $\zeta > 10^{-16}$  within their natal cloud. We fit the protocluster results with a two dimensional linear function, Equation \ref{eq:cluster_res}, showing a sub-linear trend with $\epsilon_g$ and a superlinear trend with N$_*$:
\beq
\zeta \propto \epsilon_g^{-0.24} N_*^{1.24}
\eeq
The dispersion in this relation is incredibly small due to the flatness of $\zeta(m, \mf)$. This elevated CR flux should be considered in models of protoclusters. We will explore the impact of protostellar CRs on cloud chemistry in future work.

\acknowledgements
The authors acknowledge helpful comments from an anonymous referee. SO acknowledges support from NSF AAG grant AST-1510021.

\bibliography{CRPaperI.bib}

\begin{thebibliography}{}
\expandafter\ifx\csname natexlab\endcsname\relax\def\natexlab#1{#1}\fi

\bibitem[{{Amato}(2014)}]{amato2014}
{Amato}, E. 2014, International Journal of Modern Physics D, 23, 1430013

\bibitem[{{Berezhko}(1996)}]{berezhko1996}
{Berezhko}, E.~G. 1996, Astroparticle Physics, 5, 367

\bibitem[{{Berezhko} \& {Ellison}(1999)}]{berezhko1999}
{Berezhko}, E.~G., \& {Ellison}, D.~C. 1999, \apj, 526, 385

\bibitem[{{Bricker} \& {Caffee}(2010)}]{bricker2010}
{Bricker}, G.~E., \& {Caffee}, M.~W. 2010, \apj, 725, 443

\bibitem[{{Ceccarelli} {et~al.}(2014){Ceccarelli}, {Dominik},
  {L{\'o}pez-Sepulcre}, {Kama}, {Padovani}, {Caux}, \&
  {Caselli}}]{ceccarelli2014}
{Ceccarelli}, C., {Dominik}, C., {L{\'o}pez-Sepulcre}, A., {et~al.} 2014,
  \apjl, 790, L1

\bibitem[{{Chabrier}(2005)}]{chabrier2005}
{Chabrier}, G. 2005, in Astrophysics and Space Science Library, Vol. 327, The
  Initial Mass Function 50 Years Later, ed. E.~{Corbelli}, F.~{Palla}, \&
  H.~{Zinnecker}, 41

\bibitem[{{Cleeves} {et~al.}(2013){Cleeves}, {Adams}, \&
  {Bergin}}]{cleeves2013}
{Cleeves}, L.~I., {Adams}, F.~C., \& {Bergin}, E.~A. 2013, \apj, 772, 5

\bibitem[{{Cleeves} {et~al.}(2015){Cleeves}, {Bergin}, {Qi}, {Adams}, \&
  {{\"O}berg}}]{cleeves2015}
{Cleeves}, L.~I., {Bergin}, E.~A., {Qi}, C., {Adams}, F.~C., \& {{\"O}berg},
  K.~I. 2015, \apj, 799, 204

\bibitem[{{Crimier} {et~al.}(2009){Crimier}, {Ceccarelli}, {Lefloch}, \&
  {Faure}}]{crimier2009}
{Crimier}, N., {Ceccarelli}, C., {Lefloch}, B., \& {Faure}, A. 2009, \aap, 506,
  1229

\bibitem[{{Crutcher}(2012)}]{crutcher2012}
{Crutcher}, R.~M. 2012, \araa, 50, 29

\bibitem[{{Dalgarno}(2006)}]{dalgarno2006}
{Dalgarno}, A. 2006, Proceedings of the National Academy of Science, 103, 12269

\bibitem[{{Desch} {et~al.}(2004){Desch}, {Connolly}, \&
  {Srinivasan}}]{desch2004}
{Desch}, S.~J., {Connolly}, Jr., H.~C., \& {Srinivasan}, G. 2004, \apj, 602,
  528

\bibitem[{{Drury}(1983)}]{drury1983}
{Drury}, L. 1983, \ssr, 36, 57

\bibitem[{{Favre} {et~al.}(2017){Favre}, {L{\'o}pez-Sepulcre}, {Ceccarelli},
  {Dominik}, {Caselli}, {Caux}, {Fuente}, {Kama}, {Le Bourlot}, {Lefloch},
  {Lis}, {Montmerle}, {Padovani}, \& {Vastel}}]{favre2017}
{Favre}, C., {L{\'o}pez-Sepulcre}, A., {Ceccarelli}, C., {et~al.} 2017, \aap,
  608, A82

\bibitem[{{Favre} {et~al.}(2018){Favre}, {Ceccarelli}, {L{\'o}pez-Sepulcre},
  {Fontani}, {Neri}, {Manigand}, {Kama}, {Caselli}, {Jaber Al-Edhari},
  {Kahane}, {Alves}, {Balucani}, {Bianchi}, {Caux}, {Codella}, {Dulieu},
  {Pineda}, {Sims}, \& {Theul{\'e}}}]{favre2018}
{Favre}, C., {Ceccarelli}, C., {L{\'o}pez-Sepulcre}, A., {et~al.} 2018, ArXiv
  e-prints, arXiv:1804.07825

\bibitem[{{Fontani} {et~al.}(2017){Fontani}, {Ceccarelli}, {Favre}, {Caselli},
  {Neri}, {Sims}, {Kahane}, {Alves}, {Balucani}, {Bianchi}, {Caux}, {Jaber
  Al-Edhari}, {Lopez-Sepulcre}, {Pineda}, {Bachiller}, {Bizzocchi},
  {Bottinelli}, {Chacon-Tanarro}, {Choudhury}, {Codella}, {Coutens}, {Dulieu},
  {Feng}, {Rimola}, {Hily-Blant}, {Holdship}, {Jimenez-Serra}, {Laas},
  {Lefloch}, {Oya}, {Podio}, {Pon}, {Punanova}, {Quenard}, {Sakai}, {Spezzano},
  {Taquet}, {Testi}, {Theul{\'e}}, {Ugliengo}, {Vastel}, {Vasyunin}, {Viti},
  {Yamamoto}, \& {Wiesenfeld}}]{fontani2017}
{Fontani}, F., {Ceccarelli}, C., {Favre}, C., {et~al.} 2017, \aap, 605, A57

\bibitem[{{Furlan} {et~al.}(2014){Furlan}, {Megeath}, {Osorio}, {Stutz},
  {Fischer}, {Ali}, {Stanke}, {Manoj}, {Adams}, \& {Tobin}}]{furlan2014}
{Furlan}, E., {Megeath}, S.~T., {Osorio}, M., {et~al.} 2014, \apj, 786, 26

\bibitem[{{Gaches} \& {Offner}(2018)}]{gaches2018}
{Gaches}, B.~A.~L., \& {Offner}, S.~S.~R. 2018, ArXiv e-prints,
  arXiv:1801.08555

\bibitem[{{Glassgold} \& {Langer}(1973)}]{glassgold1973}
{Glassgold}, A.~E., \& {Langer}, W.~D. 1973, \apj, 186, 859

\bibitem[{{Goldsmith}(2001)}]{goldsmith2001}
{Goldsmith}, P.~F. 2001, \apj, 557, 736

\bibitem[{{Gounelle}(2006)}]{gounelle2006a}
{Gounelle}, M. 2006, \nar, 50, 596

\bibitem[{{Gounelle} {et~al.}(2013){Gounelle}, {Chaussidon}, \&
  {Rollion-Bard}}]{gounelle2013}
{Gounelle}, M., {Chaussidon}, M., \& {Rollion-Bard}, C. 2013, \apjl, 763, L33

\bibitem[{{Grenier} {et~al.}(2015){Grenier}, {Black}, \&
  {Strong}}]{grenier2015}
{Grenier}, I.~A., {Black}, J.~H., \& {Strong}, A.~W. 2015, \araa, 53, 199

\bibitem[{{Hartmann} {et~al.}(2016){Hartmann}, {Herczeg}, \&
  {Calvet}}]{hartmann2016}
{Hartmann}, L., {Herczeg}, G., \& {Calvet}, N. 2016, \araa, 54, 135

\bibitem[{{Indriolo} {et~al.}(2010){Indriolo}, {Blake}, {Goto}, {Usuda}, {Oka},
  {Geballe}, {Fields}, \& {McCall}}]{indriolo2010}
{Indriolo}, N., {Blake}, G.~A., {Goto}, M., {et~al.} 2010, \apj, 724, 1357

\bibitem[{{Indriolo} {et~al.}(2007){Indriolo}, {Geballe}, {Oka}, \&
  {McCall}}]{indriolo2007}
{Indriolo}, N., {Geballe}, T.~R., {Oka}, T., \& {McCall}, B.~J. 2007, \apj,
  671, 1736

\bibitem[{{Indriolo} {et~al.}(2015){Indriolo}, {Neufeld}, {Gerin}, {Schilke},
  {Benz}, {Winkel}, {Menten}, {Chambers}, {Black}, {Bruderer}, {Falgarone},
  {Godard}, {Goicoechea}, {Gupta}, {Lis}, {Ossenkopf}, {Persson},
  {Sonnentrucker}, {van der Tak}, {van Dishoeck}, {Wolfire}, \&
  {Wyrowski}}]{indriolo2015}
{Indriolo}, N., {Neufeld}, D.~A., {Gerin}, M., {et~al.} 2015, \apj, 800, 40

\bibitem[{{Ingleby} {et~al.}(2013){Ingleby}, {Calvet}, {Herczeg}, {Blaty},
  {Walter}, {Ardila}, {Alexander}, {Edwards}, {Espaillat}, {Gregory},
  {Hillenbrand}, \& {Brown}}]{ingleby2013}
{Ingleby}, L., {Calvet}, N., {Herczeg}, G., {et~al.} 2013, \apj, 767, 112

\bibitem[{{Ivlev} {et~al.}(2015){Ivlev}, {Padovani}, {Galli}, \&
  {Caselli}}]{ivlev2015}
{Ivlev}, A.~V., {Padovani}, M., {Galli}, D., \& {Caselli}, P. 2015, \apj, 812,
  135

\bibitem[{{Johns-Krull}(2007)}]{johns-krull2007}
{Johns-Krull}, C.~M. 2007, \apj, 664, 975

\bibitem[{{Kim} {et~al.}(2008){Kim}, {Hirota}, {Honma}, {Kobayashi},
  {Bushimata}, {Choi}, {Imai}, {Iwadate}, {Jike}, {Kameno}, {Kameya},
  {Kamohara}, {Kan-Ya}, {Kawaguchi}, {Kuji}, {Kurayama}, {Manabe}, {Matsui},
  {Matsumoto}, {Miyaji}, {Nagayama}, {Nakagawa}, {Oh}, {Omodaka}, {Oyama},
  {Sakai}, {Sasao}, {Sato}, {Sato}, {Shibata}, {Tamura}, \&
  {Yamashita}}]{kim2008}
{Kim}, M.~K., {Hirota}, T., {Honma}, M., {et~al.} 2008, \pasj, 60, 991

\bibitem[{{Krause} {et~al.}(2015){Krause}, {Morlino}, \& {Gabici}}]{krause2015}
{Krause}, J., {Morlino}, G., \& {Gabici}, S. 2015, in International Cosmic Ray
  Conference, Vol.~34, 34th International Cosmic Ray Conference (ICRC2015), ed.
  A.~S. {Borisov}, V.~G. {Denisova}, Z.~M. {Guseva}, E.~A. {Kanevskaya}, M.~G.
  {Kogan}, A.~E. {Morozov}, V.~S. {Puchkov}, S.~E. {Pyatovsky}, G.~P.
  {Shoziyoev}, M.~D. {Smirnova}, A.~V. {Vargasov}, V.~I. {Galkin}, S.~I.
  {Nazarov}, \& R.~A. {Mukhamedshin}, 518

\bibitem[{{L{\'o}pez-Sepulcre} {et~al.}(2013){L{\'o}pez-Sepulcre}, {Taquet},
  {S{\'a}nchez-Monge}, {Ceccarelli}, {Dominik}, {Kama}, {Caux}, {Fontani},
  {Fuente}, {Ho}, {Neri}, \& {Shimajiri}}]{lopez-sepulcre2013}
{L{\'o}pez-Sepulcre}, A., {Taquet}, V., {S{\'a}nchez-Monge}, {\'A}., {et~al.}
  2013, \aap, 556, A62

\bibitem[{{Machida} {et~al.}(2007){Machida}, {Inutsuka}, \&
  {Matsumoto}}]{machida2007}
{Machida}, M.~N., {Inutsuka}, S.-i., \& {Matsumoto}, T. 2007, \apj, 670, 1198

\bibitem[{{McKee} \& {Offner}(2010)}]{mckee2010}
{McKee}, C.~F., \& {Offner}, S.~S.~R. 2010, \apj, 716, 167

\bibitem[{{McKee} \& {Tan}(2003)}]{mckee2003}
{McKee}, C.~F., \& {Tan}, J.~C. 2003, \apj, 585, 850

\bibitem[{{Offner} \& {McKee}(2011)}]{offner2011}
{Offner}, S.~S.~R., \& {McKee}, C.~F. 2011, \apj, 736, 53

\bibitem[{{Padovani} {et~al.}(2009){Padovani}, {Galli}, \&
  {Glassgold}}]{padovani2009}
{Padovani}, M., {Galli}, D., \& {Glassgold}, A.~E. 2009, \aap, 501, 619

\bibitem[{{Padovani} {et~al.}(2014){Padovani}, {Galli}, {Hennebelle},
  {Commer{\c c}on}, \& {Joos}}]{padovani2014}
{Padovani}, M., {Galli}, D., {Hennebelle}, P., {Commer{\c c}on}, B., \& {Joos},
  M. 2014, \aap, 571, A33

\bibitem[{{Padovani} {et~al.}(2013){Padovani}, {Hennebelle}, \&
  {Galli}}]{padovani2013}
{Padovani}, M., {Hennebelle}, P., \& {Galli}, D. 2013, \aap, 560, A114

\bibitem[{{Padovani} {et~al.}(2015){Padovani}, {Hennebelle}, {Marcowith}, \&
  {Ferri{\`e}re}}]{padovani2015}
{Padovani}, M., {Hennebelle}, P., {Marcowith}, A., \& {Ferri{\`e}re}, K. 2015,
  \aap, 582, L13

\bibitem[{{Padovani} {et~al.}(2018){Padovani}, {Ivlev}, {Galli}, \&
  {Caselli}}]{padovani2018}
{Padovani}, M., {Ivlev}, A.~V., {Galli}, D., \& {Caselli}, P. 2018, ArXiv
  e-prints, arXiv:1803.09348

\bibitem[{{Padovani} {et~al.}(2016){Padovani}, {Marcowith}, {Hennebelle}, \&
  {Ferri{\`e}re}}]{padovani2016}
{Padovani}, M., {Marcowith}, A., {Hennebelle}, P., \& {Ferri{\`e}re}, K. 2016,
  \aap, 590, A8

\bibitem[{{Podio} {et~al.}(2014){Podio}, {Lefloch}, {Ceccarelli}, {Codella}, \&
  {Bachiller}}]{podio2014}
{Podio}, L., {Lefloch}, B., {Ceccarelli}, C., {Codella}, C., \& {Bachiller}, R.
  2014, \aap, 565, A64

\bibitem[{{Rab} {et~al.}(2017){Rab}, {G{\"u}del}, {Padovani}, {Kamp}, {Thi},
  {Woitke}, \& {Aresu}}]{rab2017}
{Rab}, C., {G{\"u}del}, M., {Padovani}, M., {et~al.} 2017, \aap, 603, A96

\bibitem[{{Rodgers-Lee} {et~al.}(2017){Rodgers-Lee}, {Taylor}, {Ray}, \&
  {Downes}}]{rodgers-lee2017}
{Rodgers-Lee}, D., {Taylor}, A.~M., {Ray}, T.~P., \& {Downes}, T.~P. 2017,
  \mnras, 472, 26

\bibitem[{{Rudd}(1991)}]{rudd1991}
{Rudd}, M.~E. 1991, \pra, 44, 1644

\bibitem[{{Rudd} {et~al.}(1985){Rudd}, {Kim}, {Madison}, \&
  {Gallagher}}]{rudd1985}
{Rudd}, M.~E., {Kim}, Y.-K., {Madison}, D.~H., \& {Gallagher}, J.~W. 1985,
  Reviews of Modern Physics, 57, 965

\bibitem[{{Turner} \& {Drake}(2009)}]{turner2009}
{Turner}, N.~J., \& {Drake}, J.~F. 2009, \apj, 703, 2152

\bibitem[{{Umebayashi} \& {Nakano}(1981)}]{umebayashi1981}
{Umebayashi}, T., \& {Nakano}, T. 1981, \pasj, 33, 617

\bibitem[{{Watson}(1976)}]{watson1976}
{Watson}, W.~D. 1976, Reviews of Modern Physics, 48, 513

\end{thebibliography}

\appendix
\section{Cosmic Ray Spectrum Physics}\label{CRphysics}

CRs can be accelerated to near relativistic speeds in strong shocks. We use the CR model from \cite{padovani2016} and couple it to the protostellar accretion shock model described in \S\ref{methods}. Throughout this section, $\beta$ and $\gamma$ are used as proxies of energies, with $E = \gamma m_p c^2$ and $\gamma = \frac{1}{\sqrt{1 - \beta^2}}$. The CRs are assumed to be accelerated in a Bohm-type diffusion shock. The momentum distribution of the CRs from first order Fermi acceleration is
\beq
f(p) = f_0 \left ( \frac{p}{p_{\rm inj}} \right )^{-q},
\eeq
where $f_0$ is a normalization constant, $p$ is the momentum of the CR, $p_{\rm inj}$ is the injection momentum and $q$ is the power-law index. The power-law index is related to the shock compression factor, $r$, by $q = \frac{3r}{r-1}$. The distribution is defined between momenta $p_{\rm inj} < p < p_{\rm max}$, where $p_{\rm max}$ is the maximum CR momentum. The energy distribution of the shock-accelerated CRs is
\beq
\mathcal{N}(E) = 4\pi p^2 f(p) \frac{dp}{dE} ~~ ({\rm particles ~GeV^{-1} ~cm^{-3}})
\eeq
and the CR flux emerging from the shock surface is
\beq
j(E) = \frac{v(E)\mathcal{N}(E)}{4\pi} ~~ ({\rm particles ~GeV^{-1} ~cm^{-2} ~s^{-1} ~sr^{-1}}).
\eeq

The values for $p_{\rm inj}$, $p_{\rm max}$ and $r$ come from the underlying shock properties. For the compression factor, $r$, we use the hydrodynamic strong shock result
\beq
\label{eq:r}
r = \frac{(\gamma_{ad} + 1)\mathcal{M}_s^2}{(\gamma_{ad} - 1)\mathcal{M}_s^2 + 2},
\eeq
where $\gamma_{ad}$ is the adiabatic index and $\mathcal{M}_s = \frac{v_s}{c_s}$ is the sonic Mach number for the shock flow. $\mathcal{M}_s \approx 2$ at the protostellar accretion shock due to the high temperatures of the gas. The injection momentum, $p_{\rm inj}$ is related to the thermal pressure by
\beq
\label{eq:pinj}
p_{\rm inj} = \lambda p_{\rm th} = \lambda m_p c_{s,d},
\eeq
where $c_{s,d} = \frac{v_s}{r} \sqrt{\gamma_{ad}(r-1)}$ is the downstream sound speed, $m_p$ is the proton mass, and the parameter $\lambda$ depends on the shock efficiency $\eta$:
\beq
\label{eq:eta}
\eta = \frac{4}{3\sqrt{\pi}} (r - 1) \lambda^3 e^{-\lambda^2}.
\eeq 

Protostellar accretion is thought to proceed via flow along the magnetic field lines in columns connecting the disk and protostar \citep{hartmann2016}. Therefore, we assume the shock front normal is parallel to the magnetic field lines. The coefficients for upstream and downstream diffusion, $k_u$ and $k_d$ respectively, under a parallel shock are equal, $k_u = k_d$. The upstream diffusion coefficient is calculated in \cite{padovani2016}:,
\beq\label{eq:ku}
\begin{split}
k_u = \frac{2}{\tilde{P}_{CR}} \frac{V_A}{v_s} = 4 \times 10^{-2} \left ( \frac{v_s}{10^2 {\rm ~km ~s^{-1}}} \right )^{-1} \times \\
\left ( \frac{n_H}{10^6 {\rm ~cm^{-3}}} \right )^{-0.5} \left ( \frac{B}{10 {\rm ~\mu G}} \right ) \left ( \frac {\tilde{P}_{CR}}{10^{-2}} \right )^{-1}.
\end{split}
\eeq
\cite{berezhko1999} calculates $\tilde{P}_{\rm CR}$ as a function of the injection and maximum momentum. Under the approximation that $p_{\rm max} >> p_{\rm inj}$ (which is reasonable, since E$_{\rm max} \propto$ GeV and E$_{\rm inj} \propto$ keV), $\tilde{P}_{CR} = \eta r \left ( \frac{c}{U} \right )^2 \tilde{p}^a_{\rm inj} \left ( \frac{1 - \tilde{p}^b_{\rm inj}}{2r - 5} \right )$, where $\tilde{p} = p/(m_p c)$, $a = \frac{3}{r - 1}$, and $b = \frac{2r-5}{r-1}$.
The maximum energy is derived by considering different limits of the CR propagation. 

As CRs propagate through neutral gas, they undergo various kinds of interaction. CRs can excite electronic excitations and cause ionizations, with CRs with energies between 100 MeV and 1 GeV dominating the H$_2$ ionization. Furthermore, at GeV and higher energies, they can lose energy by pion production, resulting in gamma radiation. These losses are encoded in the loss function, $L(E)$. The maximum energy possible due to collisional losses is found when the acceleration rate equals the loss rate:
\beq\label{eq:eloss}
\begin{split}
\beta \left [ \frac{L(E)}{10^{-25} {\rm ~ GeV ~cm^2}} \right ] = 3.4 \frac{k_u^{\alpha}(r-1)}{r [1 + r (k_d/k_u)^{\alpha}]}  \times \\ 
\left ( \frac{v_s}{10^2 {\rm ~km ~s^{-1}}} \right )^2 \left ( \frac{n_s}{10^6 {\rm ~cm^{-3}}} \right )^{-1} \left ( \frac{B}{10 {\rm ~\mu G}} \right ),
\end{split}
\eeq
where $\beta$ (and $\gamma$) are relativistic proxies for the energy. We use the loss function $L(E)$ from \cite{padovani2009}. When neutral and ionized media are mixed, the self-generated CR wave fluctuations can be damped, decreasing the efficacy of their acceleration. The energy upper limit due to this wave damping is found by requiring that the acceleration rate is shorter than the dampening loss rate:
\beq\label{eq:edamp}
\begin{split}
\gamma \beta^2 = 8.8\times10^{-5} \tilde{\mu}^{-1} \Xi (1 - x)^{-1} \left ( \frac{v_s}{10^2 {\rm ~km ~s^{-1}}} \right )^3 \times  \\
\left ( \frac{T_s}{10^4 ~K} \right )^{-0.4} \left ( \frac{n_s}{10^6 ~{\rm cm}^{-3}} \right )^{-0.5} \left ( \frac{B}{10 ~\mu G} \right )^{-4} \left ( \frac {\tilde{P}_{CR}}{10^{-2}} \right ),
\end{split}
\eeq
where
\begin{equation*}
\Xi = \left ( \frac{B}{10 ~\mu G} \right )^4 + 1.4\times10^2 \tilde{\mu}^2 \gamma^2 \beta^2 x^2 \left ( \frac{T}{10^4 ~K} \right )^{0.8} \left ( \frac{n_s}{10^6 ~\rm{cm}^{-3}} \right )^{3}
\end{equation*}
and $\tilde{P}_{CR} = \frac{P_{CR}}{n_s m_{\rm H} U^2}$ is the fraction of the shock ram pressure that goes into the CR acceleration. CRs will diffuse out in the transverse direction of the shock. If the accretion is purely spherical, this diffusion could not happen. However, if the accretion flows along columns of gas, then this loss mechanism is taken into account. The maximum energy due to upstream escape, E$_{\rm esc,u}$, is set by requiring that the escape rate is slower than the acceleration rate:
\beq\label{eq:eescu}
\gamma \beta^2 = 4.0 \mathscr{M} k_u^{\alpha} \tilde{\mu}^{-1} \left ( \frac{v_s}{10^2 ~{\rm km ~s^{-1}}} \right ) \left (\frac{B}{10 ~\mu G} \right ),
\eeq
where $\mathscr{M} = \frac{\epsilon r_*}{10^2 {\rm AU}}$ and $\epsilon = 0.1$ \citep{berezhko1996}. The maximum CR energy, $E_{\rm max}$
\beq
\label{eq:emax}
E_{\rm max} = {\rm min}(E_{\rm loss}, E_{\rm damp}, E_{\rm esc,u})
\eeq
and the maximum momentum, $p_{\rm max}$
\beq
\label{eq:pmax}
p_{\rm max}c = \sqrt{E_{\rm max}^2 - (m_pc^2)^2}
\eeq

\section{Secondary Electron Ionizations}\label{ap:secelectrons}

Secondary electron ionizations can occur when the left over electron due to H$_2$ ionization has an energy greater than the H$_2$ ionization potential. We follow the prescription by \cite{ivlev2015} to calculate the secondary electron flux and ionization rate. The secondary electron flux is given by:
\beq
j_{\rm e}^{\rm sec}(E) = \frac{E}{L(E)} \int\limits_{E+I({\rm H_2})}^{\infty} j(E') \frac{d\sigma_p^{\rm ion}}{dE}(E, E') dE',
\eeq
where I(H$_2$) = 15.6 eV,  $L(E)$ is the collisional loss function and $\frac{d\sigma_p^{\rm ion}}{dE}$ is the proton-H$_2$ ionization differential cross section. We use the differential cross section from \cite{glassgold1973}:
\beq
\frac{d\sigma_p^{\rm ion}}{dE}(E, E') = \frac{\sigma_0(E')}{1 + \left (\frac{E'}{J} \right)^2},
\eeq
where $\sigma_0(E) = \frac{\sigma_p^{\rm ion}(E)}{J} \left [ \tan^{-1} \frac{E - I({\rm H_2})}{J} \right ]^{-1}$ is the total proton-H$_2$ ionization cross section, and J = 7 eV \citep{glassgold1973}. 
The total proton-H$_2$ ionization cross section is
\begin{align*}
\sigma_p^{\rm ion} &= ( \sigma_l^{-1} + \sigma_h^{-1})^{-1} \\
\sigma_l(E) &= 4\pi a_0^2 C x^D \\
\sigma_h(E) &= 4\pi a_0^2 [ A\ln (1+x) +B ] x^{-1},
\end{align*}
where $x = m_e E_p/m_p {\rm I(H)}$, I(H) = 13.598 eV, A = 0.71, B = 1.63, C = 0.51, and D = 1.24 \citep{rudd1985, padovani2009}. When calculating the H$_2$ ionization rate due to secondary electrons, we use the electron-H$_2$ ionization cross section: 
\begin{align*}
\sigma_e^{\rm ion} &= 8\pi a_0^2 \left ( {\rm \frac{I(H)}{I(H_2)} } \right )^2 F(t) G(t) \\
F(t) &= \frac{1 - t^{1-d}}{d - 1} \left [ \left ( \frac{2}{1+t} \right )^{d/2} \frac{1 - t^{1-d/2}}{d-2} \right ] \\
G(t) &= \frac{1}{t} \left ( A_1 \ln t + A_2 + \frac{A_3}{t} \right ),
\end{align*}
where t = $E_e/{\rm I(H_2)}$ and we adopt $d = 2.4$, $A_1 = 0.72$, $A_2 = 0.87$ and $A_3 = -0.6$ \citep{rudd1991, padovani2009}.

In principle, this process can be repeated as a cascade. However, such higher order effects will not significantly affect our results compared to other model assumptions.
\end{document}